\documentclass[conference]{IEEEtran}
\IEEEoverridecommandlockouts
\usepackage{cite}
\usepackage{amsmath,amssymb,amsfonts}
\usepackage{algorithmic}
\usepackage{graphicx}
\usepackage{textcomp}
\usepackage{xcolor}
\def\BibTeX{{\rm B\kern-.05em{\sc i\kern-.025em b}\kern-.08em
    T\kern-.1667em\lower.7ex\hbox{E}\kern-.125emX}}

\newcommand{\ignore}[1]{}%

\usepackage{multirow} 
\usepackage{comment}
\usepackage{subfig}

\usepackage{pifont}%
\newcommand{\dcircle}[1]{\ding{\numexpr181 + #1}}

\usepackage{tikz}
\usetikzlibrary{tikzmark}

\newcommand*\circled[1]{\tikz[baseline=(char.base)]{
            \node[shape=circle,draw,fill=black,inner sep=0.3pt] (char) {#1};}}
            
\newcommand*\circledwhite[1]{\tikz[baseline=(char.base)]{
            \node[shape=circle,draw,inner sep=0.3pt] (char) {#1};}}

\usepackage{color, colortbl}
\definecolor{LightCyan}{rgb}{0.8,1,1} %
\definecolor{LightGreen}{rgb}{0.5,1,0.8} %

\usepackage{makecell} %

\usepackage{amssymb} %

\usepackage{color,soul}

\begin{document}

\title{Efficient Instruction Scheduling using Real-time Load Delay Tracking}

\author{\IEEEauthorblockN{Andreas Diavastos}
\IEEEauthorblockA{
 \textit{Universitat Politècnica de Catalunya}%
}
\and
\IEEEauthorblockN{Trevor E. Carlson}
\textit{National University of Singapore}%
}

\maketitle

\pagestyle{empty}

\begin{abstract}

Many hardware structures in today's high-performance out-of-order processors do not scale in an efficient way. To address this, different solutions have been proposed that build execution schedules in an energy-efficient manner. Issue time prediction processors are one such solution that use data-flow dependencies and predefined instruction latencies to predict issue times of repeated instructions. In this work, we aim to improve their accuracy, and consequently their performance, in an energy efficient way. We accomplish this by taking advantage of two key observations. First, memory accesses often take additional time to arrive than the static, predefined access latency that is used to describe these systems. This is due to contention in the memory hierarchy and variability in DRAM access times. The use of this observed delay is important to optimize a processor's execution schedule, as previous works that use predefined information demonstrate performance losses as high as 25\%. Second, we find that these memory access delays often repeat across iterations of the same code. This, in turn, allows us to predict the arrival time of these accesses. 

In this work, we introduce a new processor microarchitecture, that replaces a complex reservation-station-based scheduler with an efficient, scalable alternative. Our proposed scheduling technique tracks real-time delays of loads to accurately predict instruction issue times, and uses a reordering mechanism to prioritize instructions based on that prediction, achieving close-to-out-of-order processor performance. To accomplish this in an energy-efficient manner we introduce: (1) an \textit{instruction delay learning mechanism} that monitors repeated load instructions and learns their latest delay, (2) an \textit{issue time predictor} that uses learned delays and data-flow dependencies to predict instruction issue times and (3) \textit{priority queues} that reorder instructions based on their issue time prediction. Together, our processor achieves 86.2\% of the performance of a traditional out-of-order processor, higher than previous efficient scheduler proposals, while still consuming 30\% less power.

\end{abstract}

\section{Introduction}
\label{sec:intro}

With each processor generation, architects aim to improve core performance while maintaining energy efficiency. To achieve high levels of performance, a processor must be able to build aggressive schedules that exploit instruction-level parallelism (ILP) and memory-level parallelism (MLP). One of the main challenges in this process is reordering instructions in a scalable, energy efficient manner. Traditional out-of-order processors schedule ready instructions using complex schedulers, that dynamically build data-flow dependencies and implicitly learn instruction delays. They achieve this by monitoring, waking up and issuing instructions once their operands are produced. However, as previous studies have shown~\cite{dynamos, mirage}, this technique uses power hungry hardware structures that inefficiently scale processor performance.

To build efficient, scalable hardware that provides both high performance and energy-efficiency, previous research has proposed a number of techniques covering both in-order and out-of-order processors. Some examples include parking non-ready instructions to better utilize available resources~\cite{ltp, Lebeck,freeway}, bypassing stalled instructions or filtering instructions based on their criticality to reduce stalling delays~\cite{lsc, casino, delayBypass, fiforder}, replaying stalled instructions to avoid blocking the instruction queue~\cite{cyclone} and instruction prescheduling using data-flow dependencies~\cite{dataflow-prescheduling,complexity-effective}. Some solutions make the realization that the schedules of general-purpose applications are highly regular and repeat during execution; these propose issue time prediction processors that try to explicitly predict when instructions will be ready to issue, using data-flow dependencies and pre-defined instruction delays~\cite{icfp, wakeup-free, gonzalez2000, gonzalez2001, widget, ooo-smt, scaling-issue-window, segmentedIQs, miss-predict}. But, unfortunately, without explicit knowledge of real-time instruction delays, the issue time predictions will never be accurate enough to achieve close-to out-of-order core performance in an efficient way. On the other hand, some works~\cite{dynamos, mirage} propose hybrid processors where an out-of-order core produces repeated instruction schedules, taking into account true memory access latency, and offloads them to simple in-order cores. However, these solutions require the implementation of two cores which increases design cost. 

In this work we aim to overcome the limitation of issue time prediction processors by dynamically building the knowledge of real-time instruction delays with low-cost hardware. In addition, we introduce an instruction reordering technique that uses this knowledge to prioritize instructions based on data-flow and timing information in a highly efficient way. We achieve high performance (and in some cases, outperform cores with expensive on-demand issue structures used by traditional out-of-order cores), with a light-weight structure that understands program dependencies and timing information to prioritize key instructions when necessary. We do this with a delay-based scheduling mechanism that uses latency information as seen by the core itself, instead of pre-defined values that have been used in all previous works up to now.

In this paper, we propose a processor microarchitecture that dynamically prioritizes the issuing of instructions, just in time for execution, by recording real-time delays of repeated loads (\textit{i.e.}, in loops) and learning data-flow dependencies of instructions to accurately predict issue times of the same instructions in future appearances. It improves energy efficiency by replacing reservation-station based instruction queues with priority queues that reorder instructions using the predicted issue time as their ordering policy, and reduces complexity by issuing only from the head of queues. 

In this work, we make the following contributions:
\begin{itemize}
    
    \item An efficient issue time prediction processor with prioritization hardware that enables instruction reordering to achieve 86.2\% of the performance of the upper-bound (a traditional out-of-order baseline), while consuming 30\% less power  (Section~\ref{sec:jit});
    \item An issue time prediction algorithm that uses real-time load delays to enable accurate prediction of issue times for repeated instructions. A prediction that facilitates the prioritization of key instructions to fill the gaps between stalled instructions, improving the performance of issue time prediction processors by 5.5\%-25\% on average (Section~\ref{sec:instructionissueprediction});
    \item A comprehensive evaluation of the proposed microarchitecture with quantitative comparison to state-of-the-art issue time prediction processors (Sections~\ref{sec:setup} and~\ref{sec:results}).
\end{itemize}

\section{Motivation and Overview}
\label{sec:overview}

One of the key reasons why out-of-order processors are able to achieve high performance is because of the aggressive scheduling of instructions. Past research suggests that out-of-order schedules are repeated~\cite{asplos2013} across loop iterations and can be learned~\cite{dynamos,mirage} or predicted by assuming a pre-defined delay, as specified in the specifications for different types of instructions~\cite{dataflow-prescheduling, miss-predict, complexity-effective, gonzalez2000, gonzalez2001}. But, although many instructions execute with static delays, like traditional addition and multiplication, load instructions that miss in the L1 can have variable latency, depending on the level of the memory hierarchy they access. 

\begin{figure}[!t]
	\centering
	\includegraphics[trim=1cm 4.5cm 1cm 4.5cm, clip=true, width=1.0 \linewidth, angle=180]{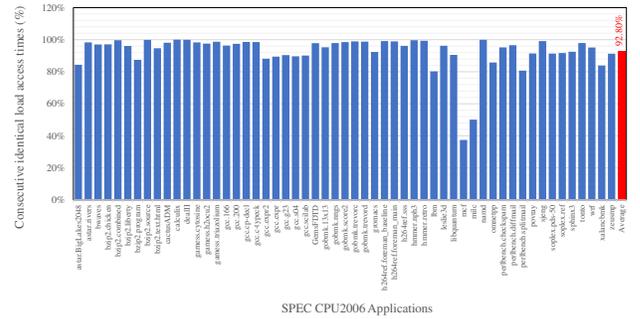}
	\caption{Percentage of repeated loads with identical access time in consecutive iterations. On average more than 92\% of loads in the SPEC CPU2006 benchmark applications have the same access time in consecutive iterations.}
	\label{fig:latency-dist}
\end{figure}

Our study shows that even for accesses to the same level of the memory hierarchy, different load instructions can have different delays due to bandwidth contention in the memory hierarchy and the variability in DRAM access times. More specifically, we see variations across all PCs of as many as 4 cycles for L2 accesses and more than 2$\times$ the number of cycles for DRAM accesses compared to the specifications defined by the implementation. Therefore, assuming a single minimum pre-defined delay for memory accesses is not sufficient to accurately predict instructions issue times. On the other hand, Figure~\ref{fig:latency-dist} shows that memory access times of loads in different appearances are repeated over consecutive iterations, on average 92.8\% of the time. 

The key insight of this paper is that \textit{to accurately predict repeated instruction issue times and build a high performance schedule, learning the latest delay of memory accesses is required}. In this work, we build schedules and reorder instructions in an energy efficient way using three core components that can replace the traditional out-of-order scheduler: (1) an instruction delay learning mechanism that tracks delays of load instructions over repeated appearances, (2) an issue time predictor that dynamically predicts when an instruction will be ready to issue and (3) priority queue reordering that use these predictions to prioritize key instructions, even after dispatch has occurred.

\section{Issue Time Prediction}
\label{sec:instructionissueprediction}

To achieve aggressive, high performing schedules we need: (1) to schedule instructions in a data-flow manner that satisfies producer-consumer relationships and (2) to reorder instructions such that idle cycles between dependent instructions are filled with independent work. Data-flow dependencies provide a scheduling policy where the execution of instructions follows the flow of the data from producers to consumers. The processor dynamically derives data-flow dependencies from the input and output operands of instructions. Because instructions require a certain amount of time to produce their output data, gaps of idle cycles are formed between dependent instructions. To achieve a high performing schedule, these gaps must be filled with independent instructions that are ready to execute. In this work, we identify these gaps by learning real-time delays of load instructions that miss in the L1 cache. All other operations have static delays; therefore, learning is not required. Assuming instructions in repeated code (e.g. loops) appear more than once, we combine instruction delays with their dependencies to predict their issue time in future appearances.

\subsection{Prediction Algorithm}

The predicted issue time of a consuming instruction (\textit{T$_{Predicted}$}(c)) is estimated as the maximum value of the addition of the predicted issue time (\textit{T$_{Predicted}$}(p)) and the delay (\textit{T$_{Delay}$(p)}) of each of its producers (Equation~\ref{issue-equation}). The delay (\textit{T$_{Delay}$(p)}) of each producer $p$ is calculated as the difference of its completion time (\textit{T$_{Complete}$}(p)) and its issue time (\textit{T$_{Issue}$}(p)) (Equation~\ref{delay-equation}). An instruction can be issued only after all its producers have completed, therefore the algorithm chooses the maximum value. By using the predicted issue time of the producers, the algorithm inherently propagates data-flow dependency chain delays to all instructions, thus the resulting predicted issue time can directly be used to order instructions.

An instruction's delay is calculated as:
\begin{equation} \label{delay-equation}
T_{Delay}(p) = T_{Complete}(p) - T_{Issue}(p)
\end{equation}

An instruction's predicted issue time is estimated as:
\begin{equation} \label{issue-equation}
T_{Predicted}(c) = \max\limits_{p=0}^{P}[T_{Predicted}(p) + T_{Delay}(p)]
\end{equation}

The proposed technique requires the core to observe and store delay information (\textit{T$_{Delay}$(p)}) of repeated instructions (remember that storing delays is only required for load instructions that miss in the L1 cache as all other instructions have static delays). In the absence of this information (first appearance of an instruction or non-repeated instructions), the algorithm assumes the lowest delay, to avoid unnecessary stalls in the execution. Because load instructions access different levels of the memory hierarchy in different iterations in an unpredictable way, constant monitoring and retraining of the predictor is required to keep the delay information up to date. Therefore, stored load delays are updated every iteration.

\subsection{Example}
\label{sec:example}

To demonstrate how the issue time prediction algorithm works, we annotate a code region, see Table~\ref{table:example}, that illustrates two loop iterations of a code snippet of the \texttt{hmmer} application from the SPEC CPU2006 benchmark suite. Although this example demonstrates reordering within one basic block, in normal execution, there are no restrictions in reordering instructions between different blocks.

For the purpose of this example, we assume a simple in-order core with 1 issue per cycle, loads/stores take 4 cycles to execute, and all other instructions execute in 1 cycle. \textit{$T_{Issue}$} corresponds to the issue cycle, \textit{$T_{Delay}$} is the instruction's delay in the current iteration and \textit{$T_{Predicted}$} is the predicted issue time for its next appearance, in relation to producers. $\Delta_{this}$ and $\Delta_{ooo}$ are the number of cycles an instruction was issued earlier, compared to a traditional in-order execution, for the proposed solution and a fully out-of-order core respectively. Note, that \textit{now} is a relative time and is different for each instruction. It merely means that an instruction is ready for execution immediately after it is dispatched.

\begin{table}[!t]
\caption{Issue time prediction example code of the \texttt{hmmer} application. We assume that instructions are already in the instruction window. $\Delta$ is the number of cycles an instruction was issued earlier in the proposed solution and an out-of-order core, compared to a traditional in-order core. Rows marked in green show the reordered instructions and in blue instructions that were issued earlier than their previous appearance. While the example uses instructions, the actual implementation uses uops.}
  \label{table:example}
  \resizebox{\linewidth}{!}{
  \begin{normalsize}
    \begin{tabular}{ll|ccccc}
      & \multirow{2}{*}{\textbf{Repeated Instructions}} & \multicolumn{5}{c}{\textbf{Prediction Algorithm}}\\[0.8ex]
      & & $T_{Issue}$ & $T_{Delay}$ & $T_{Predicted}$ & $\Delta_{this}$ & $\Delta_{ooo}$\\[.1ex]
    \hline
    
    \multirow{7}{*}{\rotatebox{90}{Iteration 1\hspace{18mm}}} & \dcircle{1} \texttt{mov (r10,rax,4),ecx} & 2 & 4 & \textit{now} & 0 & 0\\[1ex]
    & \dcircle{2} \texttt{add 0x0(r13,rax,4),ecx} & 6 & 1 & \textit{$T_{pred}^1$}+4 & 0 & 0\\[1ex] 
    & \dcircle{3} \texttt{mov ecx, 0x4(rdx)} & 7 & 4 & \textit{$T_{pred}^2$}+1 & 0 & 0\\[1ex]
    & \dcircle{4} \texttt{mov 0x18(rsp), rbx} & 8 & 4 & \textit{$now^{'}$} & 0 & -5\\[1ex]
    & \dcircle{5} \texttt{mov (r9,rax,4), r15d} & 9 & 4 & \textit{$now^{''}$} & 0 & -5\\[1ex]
    & \dcircle{6} \texttt{add (rbx,rax,4), r15d} & 13 & 1 & \textit{$T_{pred}^5$}+4 & 0 & -5\\[1ex]
    & \dcircle{7} \texttt{cmp ecx, r15d } & 14 & 1 & \textit{$T_{pred}^6$}+1 & 0 & -3\\[1ex]
    & \dcircle{8} \texttt{cmovge r15d, ecx} & 15 & 1 & \textit{$T_{pred}^7$}+1 & 0 & -3\\[1ex]
    & \dcircle{9} \texttt{mov ecx, 0x4(rdx)} & 16 & 4 & \textit{$T_{pred}^8$}+1 & 0 & -3\\[1ex]
    & ... & & & & & \\[.5ex]

    & \dcircle{1} \texttt{mov (r10,rax,4),ecx} & 20 & 4 & \textit{now} & 0 & 0\\[1ex]
    & \cellcolor{LightGreen}\dcircle{4} \texttt{mov 0x18(rsp), rbx} & \cellcolor{LightGreen}21 & \cellcolor{LightGreen}4 & \cellcolor{LightGreen}\textit{$now^{'}$} & \cellcolor{LightGreen}-5 & \cellcolor{LightGreen}-5\\[1ex]
    & \cellcolor{LightGreen}\dcircle{5} \texttt{mov (r9,rax,4), r15d} & \cellcolor{LightGreen}22 & \cellcolor{LightGreen}4 & \cellcolor{LightGreen}\textit{$now^{''}$} & \cellcolor{LightGreen}-5 & \cellcolor{LightGreen}-5\\[1ex]
    & \dcircle{2} \texttt{add 0x0(r13,rax,4),ecx} & 24 & 1 & \textit{$T_{pred}^1$}+4 & 0 & 0\\[1ex]
    & \dcircle{3} \texttt{mov ecx, 0x4(rdx)} & 25 & 4 & \textit{$T_{pred}^2$}+1 & 0 & 0\\[1ex]
    & \cellcolor{LightCyan}\dcircle{6} \texttt{add (rbx,rax,4), r15d} & \cellcolor{LightCyan}26 & \cellcolor{LightCyan}1 & \cellcolor{LightCyan}\textit{$T_{pred}^5$}+4 & \cellcolor{LightCyan}-5 & \cellcolor{LightCyan}-5\\[1ex]
    & \cellcolor{LightCyan}\dcircle{7} \texttt{cmp ecx, r15d } & \cellcolor{LightCyan}29 & \cellcolor{LightCyan}1 & \cellcolor{LightCyan}\textit{$T_{pred}^6$}+1 & \cellcolor{LightCyan}-3 & \cellcolor{LightCyan}-3\\[1ex]
    & \cellcolor{LightCyan}\dcircle{8} \texttt{cmovge r15d, ecx} & \cellcolor{LightCyan}30 & \cellcolor{LightCyan}1 & \cellcolor{LightCyan}\textit{$T_{pred}^7$}+1 & \cellcolor{LightCyan}-3 & \cellcolor{LightCyan}-3\\[1ex]
   \multirow{-9}{*}{\rotatebox{90}{Iteration 2\hspace{-18mm}}} & \cellcolor{LightCyan}\dcircle{9} \texttt{mov ecx, 0x4(rdx)} & \cellcolor{LightCyan}31 & \cellcolor{LightCyan}4 & \cellcolor{LightCyan}\textit{$T_{pred}^8$}+1 & \cellcolor{LightCyan}-3 & \cellcolor{LightCyan}-3\\[1ex]
    & ... & & & & & \\[.5ex]
    \end{tabular}
    
    \end{normalsize}
    }
\end{table}

Load instruction \dcircle{1} produces a result for instruction \dcircle{2}. Instruction \dcircle{3} is a store that depends on \dcircle{2}, while \dcircle{4} and \dcircle{5} are loads producing the operands for instruction \dcircle{6}. Instructions \dcircle{3} and \dcircle{6} are producers of instructions \dcircle{7} and \dcircle{8}, while store instruction \dcircle{9} is a consumer of instruction \dcircle{8}. Based on these dependencies, the loop consists of two major dependency chains: \dcircle{1} $\rightarrow$ \dcircle{2} $\rightarrow$ \dcircle{3} and \dcircle{4},\dcircle{5} $\rightarrow$ \dcircle{6}. These chains are independent of one another, and therefore instructions can be reordered between these chains as needed. Instruction \dcircle{2} must wait for 4 cycles before it can issue because of its dependence on load instruction \dcircle{1}. An in-order core will stall for 4 cycles between the two instructions. But an out-of-order core will fill these idle cycles by issuing instructions \dcircle{4} and \dcircle{5} earlier. To emulate this, we keep track of timing information for the relevant instructions. During the first iteration, we collect the issue cycle (\textit{$T_{Issue}$}) and the delay (\textit{$T_{Delay}$}) of every instruction and associate them with the data-flow dependencies to predict the issue time (\textit{$T_{Predicted}$}) of the same instructions in future appearances. In the second iteration, instructions \dcircle{4}, \dcircle{5} (marked in green) bypass independent instructions that have a higher predicted issue time. 

Observing the $\Delta$s in the second iteration allows us to see the benefit of this technique. After one iteration, the prediction algorithm builds a schedule that is the same as the schedule of the out-of-order core as shown by the matching deltas ($\Delta_{this}$ and $\Delta_{ooo}$). The $\Delta_{this}$ of instructions \dcircle{4} and \dcircle{5} is -5 because they are issued 5 cycles earlier compared to execution on an in-order core. Instructions that are part of the same dependency chain will also benefit, and will also be able to issue earlier (instructions \dcircle{6}, \dcircle{7}, \dcircle{8} and \dcircle{9} in the example (marked in blue)). The $\Delta_{ooo}$ is the same for both iterations because the out-of-order core can reorder instructions in every iteration.

The issue time predictor requires just one iteration to learn real time load instruction delays before applying them in the prioritization algorithm that will reorder instructions accordingly. However, in our implementation, instructions are also reordered in the first iteration by assuming L1 hit access time for all load instructions to avoid unnecessary stalls.

\section{Proposed Microarchitecture}
\label{sec:jit}

\begin{figure}[!t]
	\centering
 	\includegraphics[trim=0.5cm 20cm 0.5cm 0.5cm, clip=true, width=1.0 \linewidth]{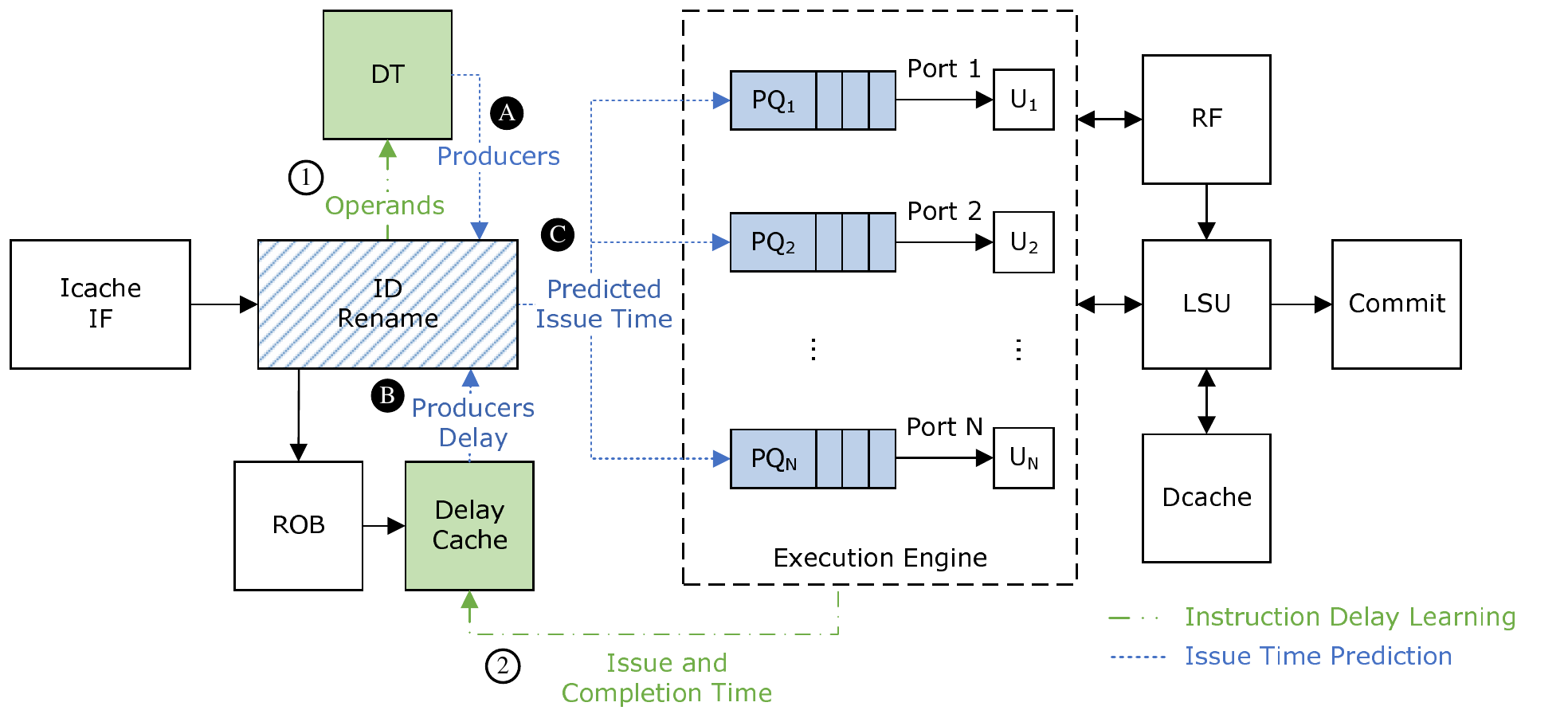}
	\caption{The proposed microarchitecture design. In green dashed lines we mark the Instruction Delay Learning process and in blue dotted lines the Issue Time Prediction process.}
	\label{fig:architecture}
\end{figure}

By tracking real-time load delays and instruction dependencies, we can more accurately predict instruction issue times and build aggressive schedules that mimic those of an out-of-order core, as shown in the example in Section~\ref{sec:example}. Using priority-based ordering hardware and the issue times predicted as the priority index it can efficiently reorder instructions. Figure~\ref{fig:architecture} shows the schematic representation of the proposed architecture. Green colored components are the structures added to implement the Instruction Delay Learning process, while blue colored structures implement the Issue Time Prediction and Priority Queue Reordering in the execution unit. 

In step \circledwhite{\scriptsize \textcolor{black}{1}} of the Instruction Delay Learning process instruction dependencies are stored in the Dependency Table (DT), that contains an entry for each physical register, and maps it to the instruction pointer that last wrote to this register. In step \circledwhite{\scriptsize \textcolor{black}{2}} the issue and completion time of load instructions that miss in the L1 cache are stored per PC in a direct-mapped memory structure called DelayCache. For every dispatched instruction the Issue Time Prediction algorithm identifies its producers from the DT in step \circled{\scriptsize \textcolor{white}{A}} and their delays from the DelayCache in step \circled{\scriptsize \textcolor{white}{B}}, to calculate the Predicted Issue Time in step \circled{\scriptsize \textcolor{white}{C}} that will be used by the Priority Queues (PQs) as a priority index to reorder instructions in the execution engine. %

\subsection{Instruction Delay Learning}

While instructions are being fetched, decoded and renamed, dependencies are stored and built by the Dependency Table (DT). As instructions start executing, delays of instructions that caused upcoming instructions to stall (L1 cache miss) are stored in the DelayCache, initiating the training of the prediction mechanism. In this work, the delay represents the execution time of an instruction with respect to its issue time. An alternative approach not used in our final design stores the delay of an instruction with respect to its dispatch time, but our study shows large slowdowns in such design due to the unpredictability of structural hazards (see Figure~\ref{fig:dispatch-time}).

Although most instructions require a static delay before delivering their result, it is loads that cause the majority of the stalls, and their delay can be variable and unknown at dispatch time. In this implementation we only store delays of loads that miss in the L1 and for all other instructions we use their predefined delay (derived from their type or L1 access time for loads that don't miss). This allows us to minimize the storage overhead and power requirements when implementing the DelayCache. 

Due to application characteristics that relate to branch behavior and memory access patterns, load delays are unpredictable in different iterations (see Section~\ref{subsec:analysis}/Figure~\ref{fig:training}). Therefore, the DelayCache is continuously updated with the latest delay for every stored instruction and the issue time predictor is trained in every iteration of a repeated code. Our experiments show that, for the majority of the applications tested, training every iteration produces the highest performance. Activity-based power analysis shows that training is not expensive, as only a subset of load instructions have a variable delay that require an update to the DelayCache.

Although our implementation trains the issue time predictor using load delay information, the issue prediction mechanism can be applied as-is to other instructions with variable delay, such as floating point division and transcendental functions. In this work, we do not cover their potential performance benefits, as they do not occur often in the applications we evaluate.

\subsection{Issue Time Prediction and Dispatch}
\label{sec:steering}

The delays of instructions stored in the DelayCache combined with the dependencies from the Dependency Table (DT), provide the necessary inputs for predicting the issue time of instructions as described in Section~\ref{sec:instructionissueprediction}. For every renamed instruction, the DT is queried using the instruction's input operands to find possible producers. In a DT hit, the DelayCache is queried with the producer's addresses, and correspondingly, in a DelayCache hit, the delay will be retrieved to calculate the current instruction's issue time. In case of a miss in the DelayCache, the value of an L1 hit (4 cycles in our microarchitecture) is used to avoid unnecessary delays in the absence of misses.

The Execution Engine, which has the primary task of reordering instructions, is built using multiple priority instruction queues, with each functional unit having its own dedicated queue. Although instructions from multiple queues can execute out-of-order, instructions in a single queue can be issued only from the head of the queue and only to the corresponding functional unit. Because each queue corresponds to a specific functional unit, instructions are dispatched to the queues according to their type. If an instruction matches to more than one queue, data-flow dependencies are used to steer incoming instructions to the first queue that has a producing instruction at its tail, otherwise it will go to the queue with the least number of instructions. Our studies indicate that round-robin and global dependence steering schemes reduce performance compared to our scheduling methodology (see Figure~\ref{fig:steering} for more details).

Issue times are predicted for first time appearing or non-repeated instructions even in the absence of delay information by assuming the lowest delay (L1 access hit), to avoid unnecessary execution stalls.

\subsection{Priority Queue Reordering and Issue}
\label{sec:microarch:issue}

In the proposed architecture we remove the traditional reservation-station-based scheduler, and instead, reorders instructions using light-weight and efficient priority instruction queues in the execution engine. Priority Queues (PQ) are built using Systolic Priority Queues~\cite{leiserson1979systolic}, where instructions are reordered using a priority index (their predicted issue time in this case). Insertion and removal in a priority queue happen at the head as described in~\cite{leiserson1979systolic} thus, highest-priority inserted instructions are directly available from the head of a PQ on the next cycle, therefore back-to-back instruction execution is achieved. A free list of entries in the queues is also used so that new entries can be inserted at a free position. Because each functional unit has its own instruction queue and only instructions at the head of each queue can be issued, complex selection logic is not required to decide which instructions to issue every cycle. When an instruction at the head of a queue has unresolved data dependencies the queue blocks. However, instructions in other queues are not affected as only instructions in a blocked queue will stall. %

\subsection{Register Renaming}

Register renaming works in the same way as in traditional out-of-order processors. Renaming replaces destination architectural registers with physical registers to eliminate the name dependencies (output dependencies and anti-dependencies) between instructions and it automatically recognizes true dependencies. True data dependencies between instructions allow for a more flexible execution of instructions. Maintaining the status for each register, indicating whether or not it has been computed yet, allows the execution of instructions to be performed out-of-order when there are no true data dependencies.

\subsection{Memory Dependencies}

Memory operations are also reordered to maximize performance. Contrary to register dependencies that can be resolved at decode time, store-to-load memory dependencies with overlapping memory addresses can lead to incorrect execution if loads or stores are executed before older stores that refer to the same address. Memory dependencies are accurately predicted by identifying the stores upon which a load depends (store set), and communicate that information to the issue time predictor~\cite{store-set-predictor}. Similarly to a traditional out-of-order processor, using the ROB and the LSU prevents memory violations. The LSU tracks executing memory operations and makes sure that they are committed in program order. Instructions are verified before commit to ensure that no memory violations will be visible to the architecture state.

\subsection{Commit}

The commit stage checks for exceptions before it releases structures such as store buffer entries and rename registers. Instructions enter in-order into the ROB during dispatch, record their completion out-of-order, and leave the ROB in-order. Interrupts and branch misspeculation events are handled as in other conventional processors. However, retraining of the issue time predictor is not required in this case and if the core matches a repeated instruction from the DelayCache, it will be reordered immediately.

\subsection{Multi-core Support}

In a multi-core implementation, new connections are added in the memory hierarchy for loads accessing remote memory locations. Issue time prediction in the proposed design is based on memory access latency at any part of the memory hierarchy; therefore, the prediction algorithm will adapt accordingly and learn remote access delays. As the coherence misses could be less predictable, it would require new studies and, potentially, structure changes to handle these cases. However, this is out of the scope of this work. This core, as implemented, does not change any significant components in the back-end of the processor and, therefore, is compatible with the original coherence and consistency models as described in the core.

\section{Experimental Setup}
\label{sec:setup}

\begin{figure*}[ht]
	\centering
	\includegraphics[trim=1.85cm 11.8cm 1.95cm 11.8cm, clip=true, width=0.9 \linewidth]{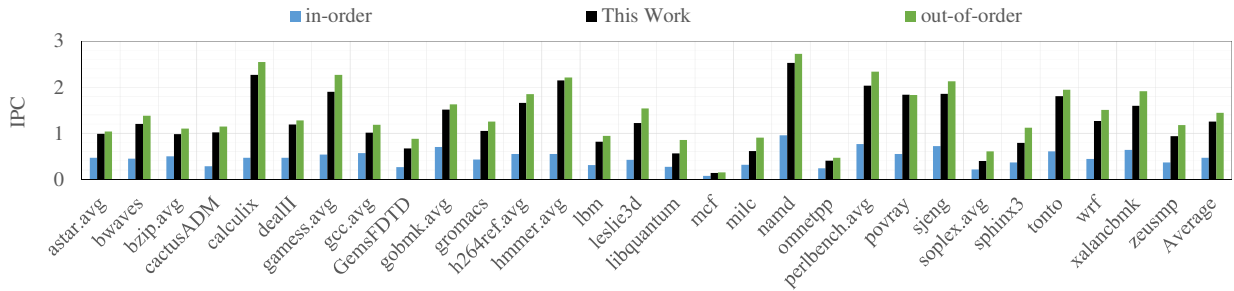}
	\caption{Performance of the proposed implementation compared to in-order and out-of-order baseline processors. For clarity, we plot average values for applications with multiple inputs (\textit{\textless application\_name\textgreater.avg}).}
	\label{fig:performance}
\end{figure*}

The performance evaluation of this work was performed on a modified version of the Sniper Multi-Core Simulator~\cite{sniper}, version 6.2 that uses the Instruction Window-Centric core model~\cite{sniper-taco}. We use a detailed DRAM model that takes into account DRAM page locality, and other low-level details that account for all detailed DRAM delays. Power and energy analysis was conducted with McPAT~\cite{mcpat} version 1.3, modified to support our microarchitecture. Applications were compiled with the GCC compiler (-O2 optimization flag) and executed with the reference inputs of the SPEC CPU2006 benchmarks, using a single, representative (SimPoint-based~\cite{simpoint}), 750 million instruction trace. Average results are computed by combining output results of common workloads (but different input) into a weighted value before averaging the results across applications. The details of added structures to the core, with area and average power consumption, are listed in Table~\ref{table:addons} and the details of the simulated microarchitectures are listed in Table~\ref{table:setup}. Performance is measured in Instructions per Cycle (IPC) and energy efficiency in Million Instructions Per Second per Watt (MIPS/Watt) and Energy Delay Product (EDP). Unless explicitly stated, all summary results are weighted average values of all applications, while black bars represent results of the proposed design configuration described in Table~\ref{table:setup}.

\begin{table}[!t]
  \caption{Power and area of the new design structures. In parenthesis is their overhead over the entire core. The Priority Queues are implemented using 2$\times$13 entries per unit to match the 64 entries of the out-of-order baseline.}
  \label{table:addons}
  \centering
  \scalebox{0.8}{
    \begin{tabular}{llcrr}
        \hline
        Component & Organization & Ports & Area ($\mu$m$^2$) & Power (mW) \\[.1ex]
        \hline
        DT & 256 entries $\times$ 1B & 12r4w & 14.54 (0.37\%) & 11.74 (0.37\%) \\[.1ex]
        DelayCache & 512 entries $\times$ 12B & 4r1w & 103.31 (3.25\%) & 43.41 (1.87\%) \\[.1ex]
        Priority Queues & 5 $\times$ 2 $\times$ 13 entries $\times$ 1B & 1r1w & 0.28 (0.01\%) & 1.49 (0.05\%) \\[.1ex]
        \hline
    \end{tabular}
    }
\end{table}

\begin{table}[!t]
  \caption{Simulated microarchitecture parameters.}
  \label{table:setup}
  \centering
  \scalebox{0.95}{
    \begin{tabular}{|l|ccc|}
      \hline
        \textbf{Component} &  \multicolumn{3}{c|}{\textbf{Parameters}}\\[0.5ex]
            & in-order & This Work & out-of-order \\[0.5ex]
        \hline
        Core              & \multicolumn{3}{c|}{2GHz, superscalar}  \\
        Issue width       & 4-way & 4-way & 4-way  \\
        Reorder logic     & none & 128-entry ROB & 128-entry ROB,\\
                          &  &  5$\times$13-entry PQs & 64-entry RS\\
        DT                & -   & 256 entries ($\times$ 1B) &  -\\
        DelayCache        & -   & 512 entries ($\times$ 12B)  & -\\
        Branch Predictor  & \multicolumn{3}{c|}{TAGE-SC-L~\cite{seznec:hal-01086920}}  \\
        Branch Penalty    & 6 cycles & 8 cycles & 8 cycles \\
        Execution units   & \multicolumn{3}{c|}{2 int, 1 fp, 1 branch, 1 load/store} \\
        L1-I Cache        & \multicolumn{3}{c|}{32KB, 4-way LRU} \\
        L1-D Cache        & \multicolumn{3}{c|}{32KB, 8-way, LRU, 4 cycle, 8 outstanding} \\
        L2 cache          & \multicolumn{3}{c|}{512KB, 8-way, LRU, 8 cycle, 12 outstanding} \\
        Prefetcher        & \multicolumn{3}{c|}{L1, stride-based, 16 independent streams} \\
        Main memory       & \multicolumn{3}{c|}{DDR3-1600, 800 MHz, ranks: 4, banks: 8,}\\
                          & \multicolumn{3}{c|}{page size: 4KB, bus: 64 bits,}\\
                          & \multicolumn{3}{c|}{\texttt{tRP-tCL-tRCD:} 11-11-11}  \\
        Technology node   & \multicolumn{3}{c|}{28nm} \\
      \hline
    \end{tabular}
    }
\end{table}

\section{Results and Analysis}
\label{sec:results}

\subsection{Performance Analysis}
\label{subsec:performance}

The proposed processor achieves 2.7$\times$ and 86.2\% of the performance of the baseline in-order and out-of-order cores respectively (Figure~\ref{fig:performance}). Although instructions issue only from the head of the instruction queues, it achieves near-out-of-order performance by de-prioritizing instructions that were predicted to stall the execution (\textit{i.e.} consumers of loads that do not hit in the L1 cache). This allows ready instructions to move to the head of the instruction queues. Note that when an instruction at the head is not ready to issue, the queue will block.

Per instruction analysis shows that loads and their address generating instructions are issued earlier in the new design, compared to the out-of-order baseline. This happens because address generating instructions rarely depend on long-latency operations~\cite{lsc}, therefore, the new processor predicts shorter issue times for them and their consuming loads, even compared to older instructions that are also ready to issue. In an age-based ordering scheduler of an out-of-order core however, ready instructions are issued based on their fetched order. Therefore, loads that are issued earlier result in shorter data waiting time. This is reflected in applications, like \texttt{astar}, \texttt{dealII} and \texttt{povray} where this work's performance meets or exceeds the performance of the out-of-order. While in general, this work performs as well as the out-of-order for compute-intensive applications, there are a few that show lower performance. Applications like \texttt{gamess} that are not bound by long-latency memory accesses, stress the multi-queue backend of the new design, where instructions can issue from the head of a queue, to the corresponding functional unit only. 

The main reasons the proposed processor is unable to meet the performance of the out-of-order processor are: (1) the per functional unit instruction queue design, (2) the prediction algorithm training that requires at least one iteration to learn real time load delays and (3) the accuracy of using the previous load delay to predict the next delay. However, the design is a trade-off made to significantly improve the processor's overall energy efficiency. An alternative single in-order issue queue would severely limit the performance, while adding selection logic over queues that can issue to multiple units or using reservation-station-based queues would greatly increase power consumption (see Figure~\ref{fig:vliw}). While the delay prediction training and accuracy is an application dependent overhead that does not have a major impact on overall performance (see Figure~\ref{fig:training}), even on a larger core. Our analysis shows that on a scaled-up, Skylake-like processor, the additional overhead is only 3\%.

\subsection{Power and Efficiency Analysis}
\label{subsec:efficiency}

\begin{figure}[!t]
  \centering
  \subfloat[\textbf{Power consumption}]{\includegraphics[trim=2cm 7.2cm 1.5cm 7.5cm, clip=true, width=0.5 \linewidth]{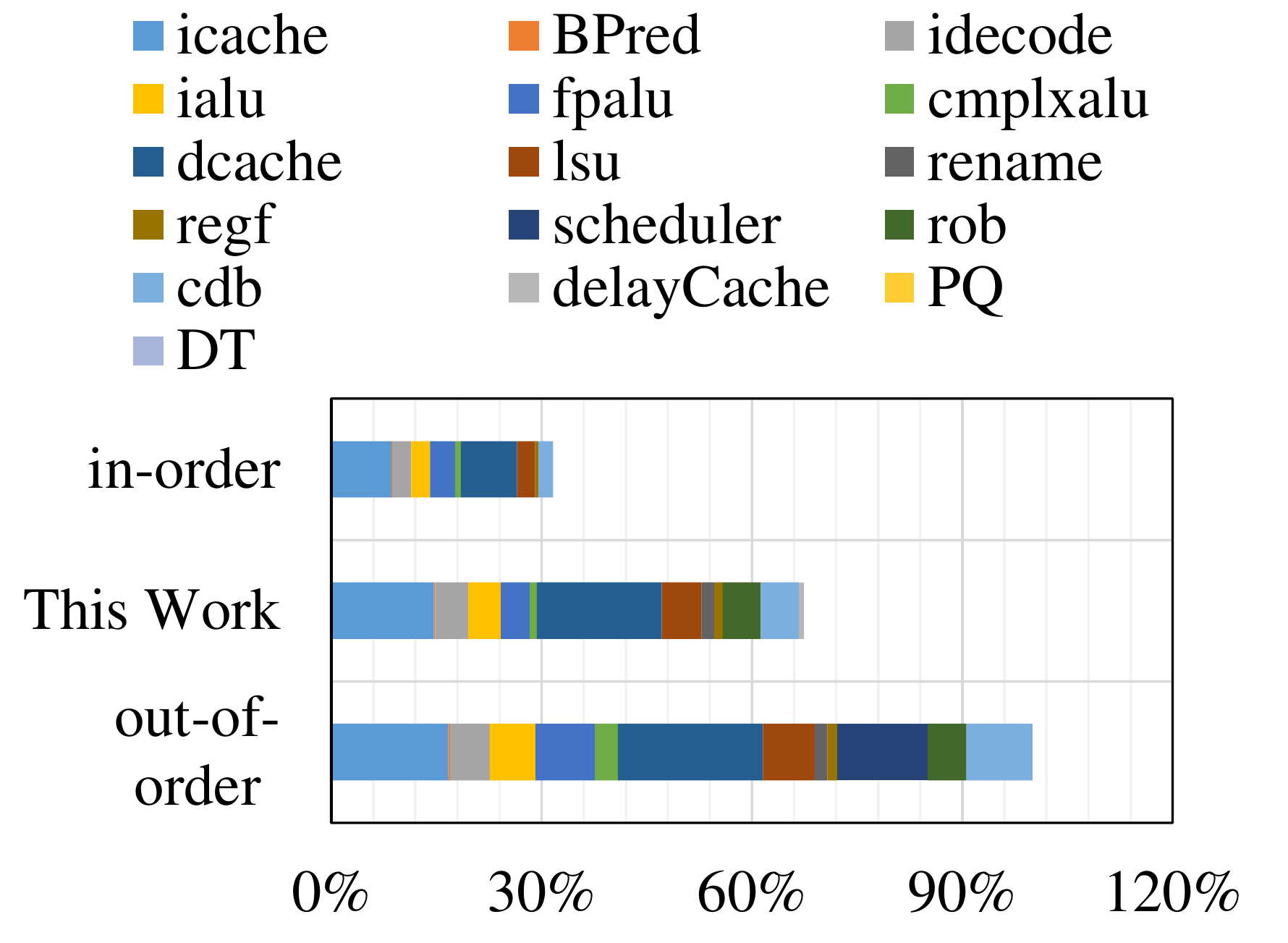}\label{fig:power}}
  \hfill
  \subfloat[\textbf{Energy efficiency}]{\includegraphics[trim=2cm 7.2cm 1.5cm 7.5cm, clip=true, width=0.5 \linewidth]{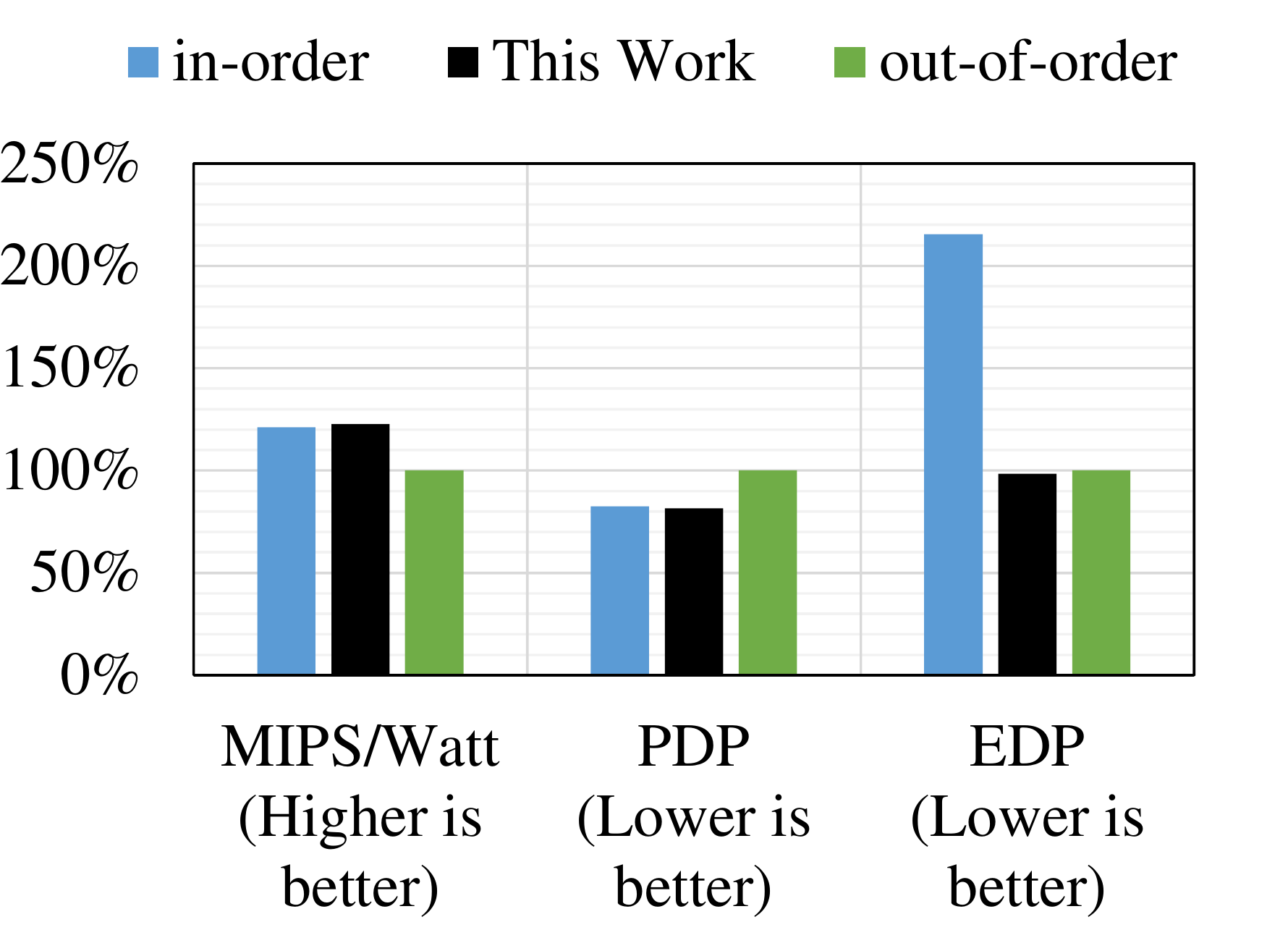}\label{fig:efficiency}}
  \caption{Normalized to the out-of-order: (a) Power consumption, (b) Efficiency (MIPS/Watt), Power Delay Product (PDP) and Energy Delay Product (EDP).}
\end{figure}

Figure~\ref{fig:power} shows power results for the same processors, normalized to the out-of-order baseline. The in-order core consumes 31.6\% the power of the out-of-order, while the proposed processor consumes 67.4\% its power. One of the main reasons for the power reduction in this work is the removal of the reservation-station-based instruction scheduler that takes 13\% of the total power of the out-of-order core (including wake-up and selection logic). The rest of the power gained is coming from the difference in runtime compared to the out-of-order. As performance increases, the amount of dynamic power also increases. Dynamic power is data dependent and is closely tied to the number of transistors that change state~\cite{power}. The DelayCache, the Priority Queues and the DT contribute only 2\% to the total power of the new core. Priority queues are efficiently implemented using simple interconnected FIFO queues, while the small number of delays that need to be stored allows for a small size DelayCache with few accesses, consequently little dynamic power consumed.

Figure~\ref{fig:efficiency} outlines the energy efficiency normalized to the out-of-order core. Despite its low performance, the simplicity and low-power hardware of the in-order core provide a 21\% increase in efficiency over the more complex out-of-order core. The significantly higher performance of this work, in conjunction with the lower total power, achieves an improvement of 22.7\% over the out-of-order. On the right side of Figure~\ref{fig:efficiency}, efficiency is outlined as a metric of the Power Delay Product normalized to the out-of-order core. The proposed processor achieves a reduction of 19\% and 1\% in PDP compared to the out-of-order and in-order respectively.

\subsection{State-of-the-art Issue Time Predictors}
\label{subsec:actual-related-work}

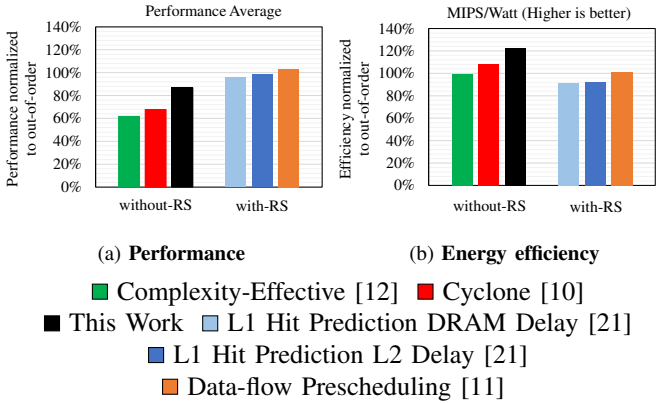
\begin{figure}[!t]
  \centering
  \subfloat[\textbf{Performance}]{\includegraphics[trim=2cm 7.2cm 1.5cm 8.3cm, clip=true, width=0.5 \linewidth]{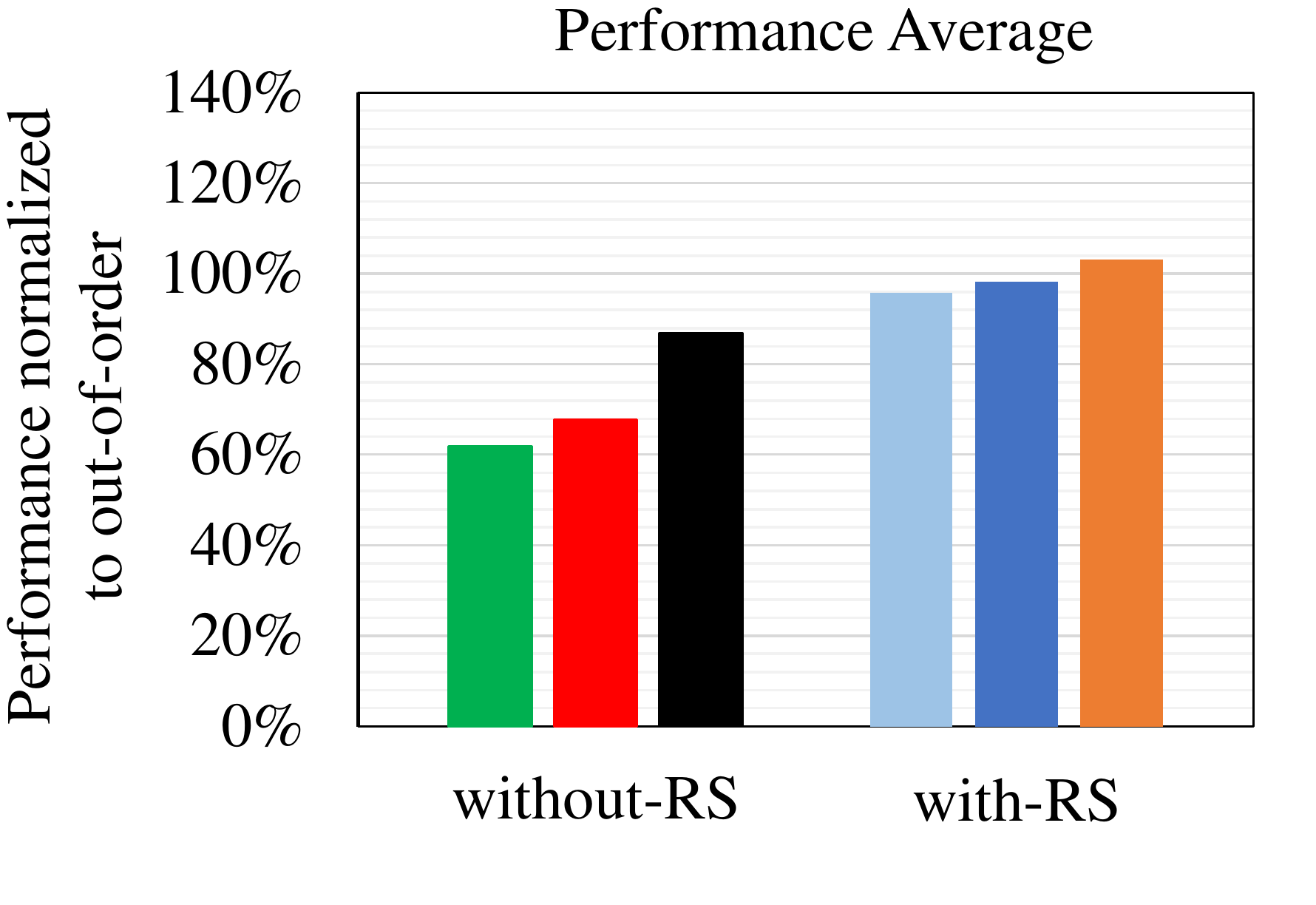}\label{fig:actual-related-work-performance}}
  \hfill
  \subfloat[\textbf{Energy efficiency}]{\includegraphics[trim=2cm 7.2cm 1.5cm 8.3cm, clip=true, width=0.5 \linewidth]{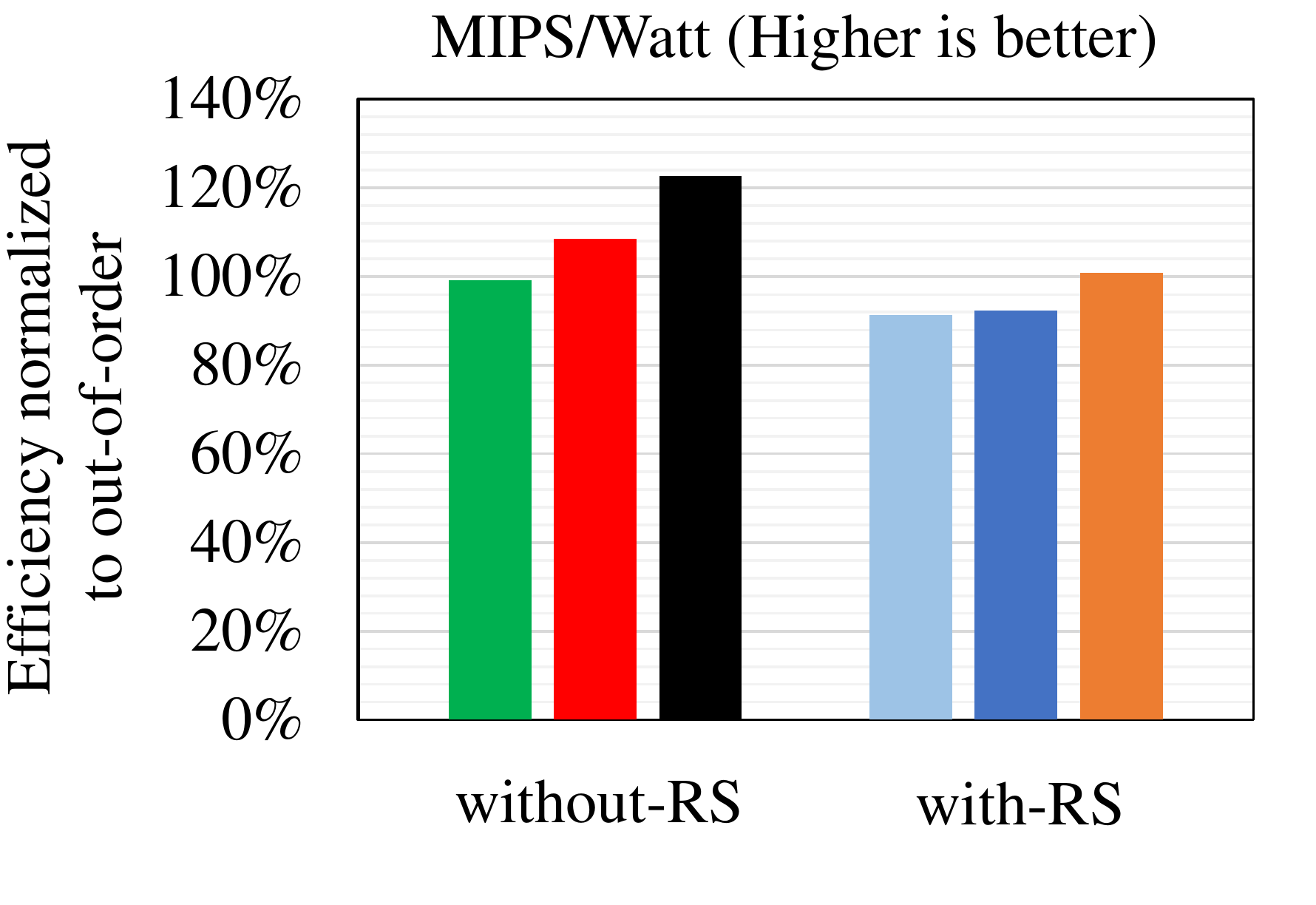}\label{fig:actual-related-work-efficiency}}
  
    \definecolor{1}{RGB}{0,176,80}
    \definecolor{2}{RGB}{255,0,0}
    \definecolor{3}{RGB}{0,0,0}
    \definecolor{4}{RGB}{157,195,230}
    \definecolor{5}{RGB}{68,114,196}
    \definecolor{6}{RGB}{237,125,49}
    \vspace{2 mm}
    \tikz \draw [fill=1] (0,0) rectangle (0.2 ,0.2); Complexity-Effective [12]~
    \tikz \draw [fill=2] (0,0) rectangle (0.2 ,0.2); Cyclone [10]~\\
    \tikz \draw [fill=3] (0,0) rectangle (0.2 ,0.2); This Work~
    \tikz \draw [fill=4] (0,0) rectangle (0.2 ,0.2); L1 Hit Prediction DRAM Delay [21]~\\
    \tikz \draw [fill=5] (0,0) rectangle (0.2 ,0.2); L1 Hit Prediction L2 Delay [21]~\\
    \tikz \draw [fill=6] (0,0) rectangle (0.2 ,0.2); Data-flow Prescheduling [11]~
    
  \caption{(a) Performance and (b) Energy efficiency of state-of-the-art issue time prediction processors, normalized to the baseline out-of-order processor.}
  \label{fig:actual-related-work}
\end{figure}

Figure~\ref{fig:actual-related-work} shows performance and energy efficiency results of state-of-the-art issue time predictors implemented on top of our baseline processors and compared to this work. We categorize these processors to those that eliminate the traditional reservation-station-based scheduler (\textit{without-RS}) and those that still use it (\textit{with-RS}). To calculate the efficiency for the processors \textit{with-RS} we used the power consumption of the out-of-order core as-is, without adding the overheads of their added structures. We use these results only as a reference and note that in a real implementation, their efficiency would actually be lower. 

The \textit{Complexity-Effective}~\cite{complexity-effective} solution steers instructions to in-order queues based on their dependencies alone. Dependent instructions are steered to the same queue, while independent instructions are steered to empty queues. This solution only achieves 61.8\% of the out-of-order core performance because it does not take advantage of the delays between dependent instructions. \textit{Cyclone}~\cite{cyclone} uses dependencies to predict instruction issue times and employs a selective replay mechanism for stalled instructions. However, performance is low due to the conflicts arising during instruction flow because of its queue structure and the limitation that only instructions at the head of the queue are candidates for issuing~\cite{wakeup-free}. Using real-time delay information on top of data-flow dependencies to predict instruction issue times in \textit{This Work}, solves these problems and achieves significant performance improvement over other processors \textit{without-RS}.

Predicting only L1 hits~\cite{miss-predict} and assuming L2 (\textit{L1 Hit Prediction L2 delay}) or DRAM (\textit{L1 Hit Prediction DRAM delay}) access delays for all other loads does not improve out-of-order processor performance, because it ignores the actual miss delay that is the key factor for stalling the pipeline. Assuming L2 delay for all misses ignores DRAM accesses and stalls the pipeline for extended periods of time and using DRAM delay makes dependent instructions wait for an unnecessary amount of time, even though they are ready to execute. \textit{Data-flow Prescheduling}~\cite{dataflow-prescheduling} reorders instructions before sending them to the instruction window using data-flow dependencies and assuming a L1 cache-hit delay for all loads. This optimistic assumption improves performance over the out-of-order core as it does not delay ready instructions in the issue window. However, as in all solutions \textit{with-RS}, misspredicted instructions will not stall the pipeline because they will be overlapped using the out-of-order scheduler.

In general, processors \textit{with-RS} produce higher performance (Figure~\ref{fig:actual-related-work-performance}) because the reordering is handled by their reservation-station-based instruction queue. However, processors \textit{without-RS} achieve higher energy efficiency (Figure~\ref{fig:actual-related-work-efficiency}) because of the simplicity of their design. These results highlight the importance of using real-time delay information to provide out-of-order performance when predicting instruction issue times, while reordering instructions using priority queues will achieve it in an energy efficient way (as we demonstrate with \textit{This Work}). We note that previous solutions investigated their effectiveness using very large cores. We performed experiments using similar simulation configurations and this work scales in a similar way. %

\subsection{State-of-the-art Issue Time Predictors on the Proposed Hardware}
\label{subsec:against-related-work}

\begin{figure}[!t]
	\centering
	\includegraphics[trim=1cm 8.5cm 1cm 8.5cm, clip=true, width=1.0 \linewidth]{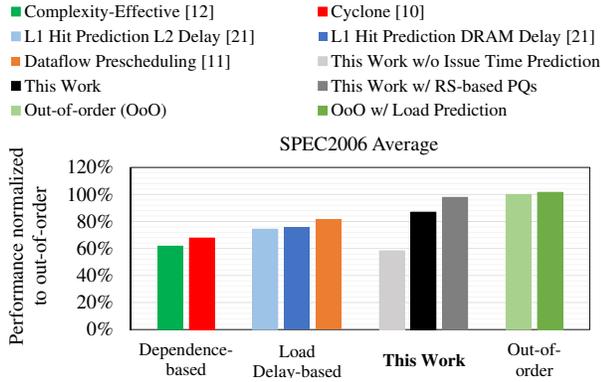}
	\caption{Performance of state-of-the-art issue time prediction techniques implemented on the proposed microarchitecture, normalized to the out-of-order baseline. Because Complexity-Effective does not use a prediction mechanism, its microarchitecture is implemented precisely as described in~\cite{complexity-effective}.}
	\label{fig:load-delay-impact}
\end{figure}

In Figure~\ref{fig:load-delay-impact} we show results for the same state-of-the-art issue time prediction techniques implemented on-top of the proposed microarchitecture. With this study we highlight the importance of using real-time delays and their effectiveness in predicting instruction issue time. We categorize these solutions to those that use only dependencies to reorder instructions (Dependence-based) and those that use both dependencies and load delays (Load Delay-based).

Dependence-based solutions achieve low performance because using only dependencies between instructions does not take into account the idle time between dependent instructions. Load Delay-based solutions outperform Dependence-based solutions, but using a static delay, like \textit{Data-flow Prescheduling}~\cite{dataflow-prescheduling}, for all types of instructions results in an average performance loss of 5.5\% compared to the \textit{This Work} (detailed analysis shows up to 32\% performance loss for memory-intensive applications). Predicting L1 hits~\cite{miss-predict} and assuming L2 (\textit{L1 Hit Prediction L2 delay}) or DRAM (\textit{L1 Hit Prediction DRAM delay}) delays also incurs significant overhead as they ignore access time to other memory regions and stall instructions at the heads of the in-order multi-queue backend of this design. %

Using the \textit{instruction delay learning} mechanism on the baseline out-of-order improves performance only by 1.6\%. The ability of reservation-station-based scheduler to monitor and issue instructions based on operand availability is sufficient enough to get optimal performance. However, applying these techniques on top of a simpler in-order-based core offers large performance benefits as results for the \textit{This Work} illustrate. A limit study (\textit{This Work w/ RS-based PQs}) can achieve 97.9\% of the out-of-order core performance when associative lookups are performed in the PQs to minimize issue time mispredictions.

\subsection{Proposed design Implementation Analysis}
\label{subsec:analysis}

\begin{figure}[!t]
  \centering
  \subfloat[\textbf{Debut vs Repeated}]{\includegraphics[trim=2cm 7.8cm 1.5cm 9.5cm, clip=true, width=0.5 \linewidth]{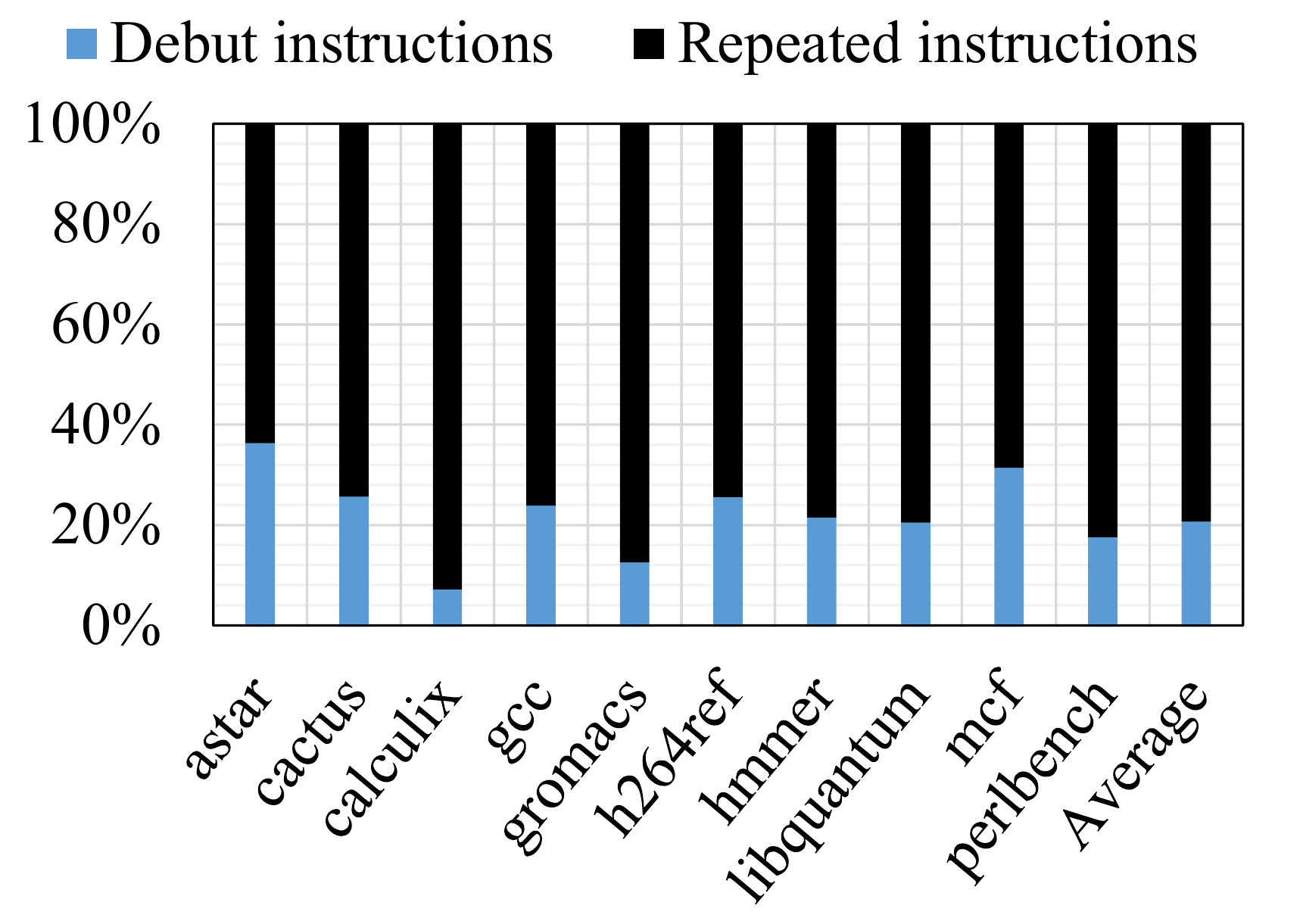}\label{fig:repeated-code}}
  \hfill
  \subfloat[\textbf{Dependency stalls}]{\includegraphics[trim=2cm 7.8cm 1.5cm 7.9cm, clip=true, width=0.5 \linewidth]{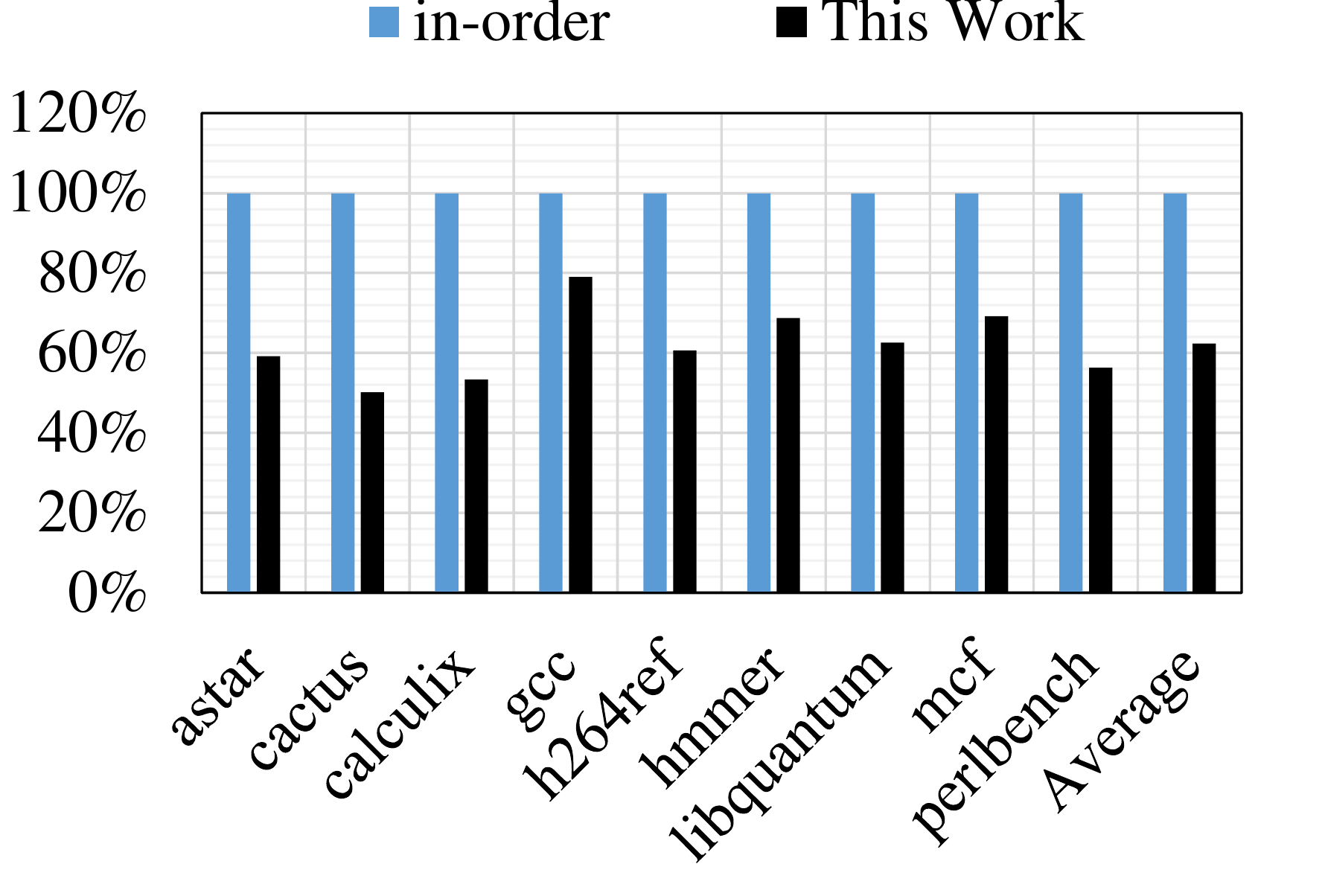}\label{fig:d-stalls}}
  \caption{(a) Cycles spent executing \textit{Debut} instructions (appear for the first time or miss in the DelayCache) and \textit{Repeated} instructions (hit in the DelayCache), (b) Cycles the instruction queues are blocked due to unresolved dependencies.}
\end{figure}

In this work we take advantage of the high levels of repeatably of the code~\cite{asplos2013} to learn the delays of instructions up-front and prioritize them on future encounters. Figure~\ref{fig:repeated-code} shows the number of cycles each application spends executing Repeated and Debut instructions. Instructions that appear more than once during execution are called Repeated, while instructions seen for the first time are called Debut (including the first appearance of Repeated instructions). Some applications (like \texttt{astar} and \texttt{mcf}) see as much as 36\% Debut instructions. While some of this is an artifact of application sampling (see Section~\ref{sec:setup}), there will always exist code that is seen only once, either because of large Debut code or because of large number of loops that do not fit in the DelayCache for the entire execution of the application. Overall, a large number of Debut instructions can potentially reduce performance as the issue time predictor will not have real-time information for these instructions and will have to use static delays instead, that can produce issue time mispredictions.

In Figure~\ref{fig:d-stalls}, we show the number of stalls at the head of the queues due to unresolved dependencies, normalized to the in-order processor baseline. The out-of-order processor is not represented in this figure as it allows issuing from any position in the instruction queue. Some applications, like \texttt{gcc}, have long dependency chains and stall the processor more often, while compute-intensive applications, like \texttt{cactus}, can expose more ILP. Overall, the proposed core reduces stalled cycles at the head of instruction queues by an average of 38\%.

\begin{figure}[!t]
  \centering
  \subfloat[\textbf{Prediction using instruction dispatch time}]{\includegraphics[trim=2cm 6.8cm 1.5cm 8.3cm, clip=true, width=0.48 \linewidth]{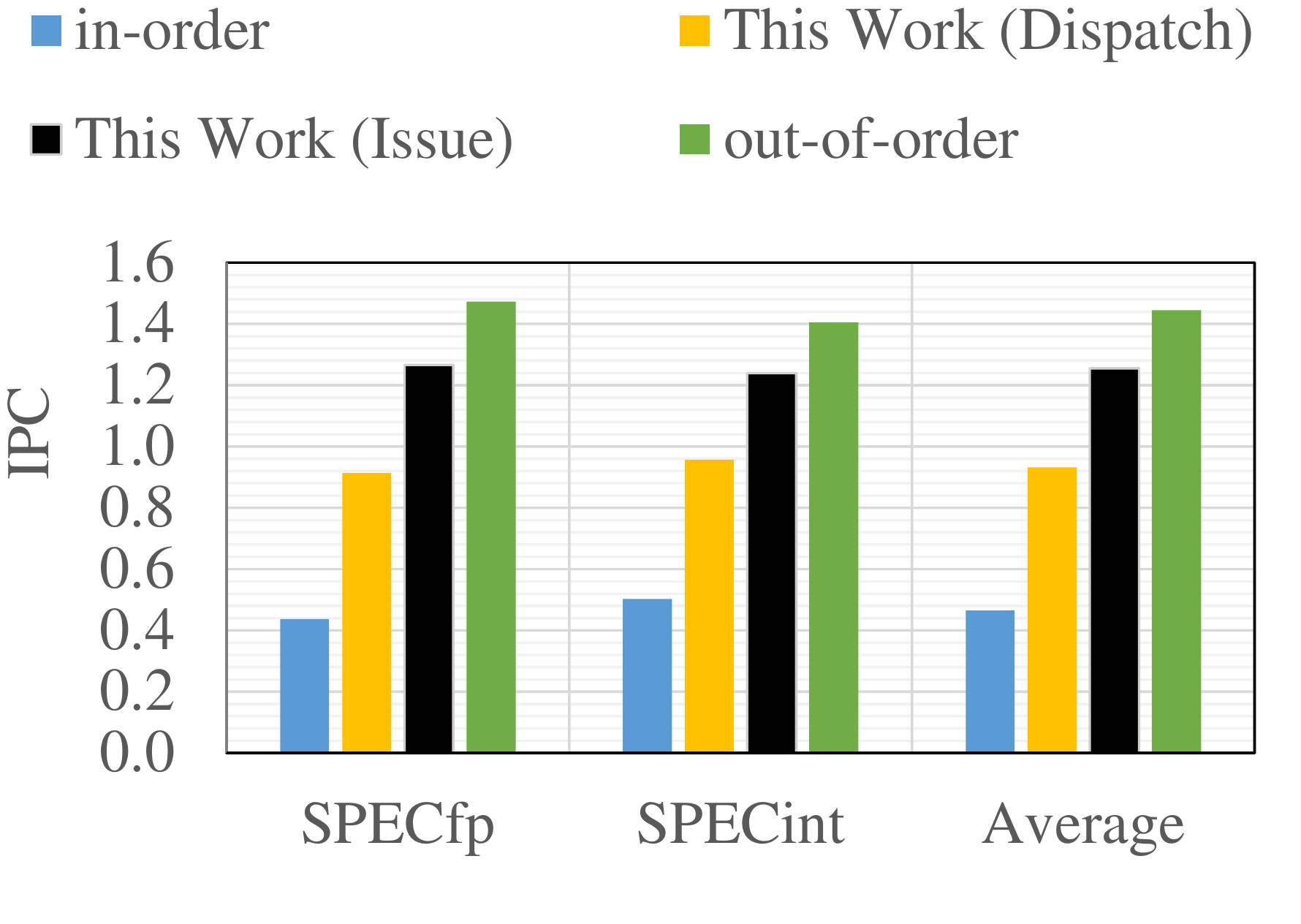}\label{fig:dispatch-time}}
  \hfill
  \subfloat[\textbf{Execution engine implementations for all cores}]{\includegraphics[trim=2cm 6.8cm 1.5cm 7.5cm, clip=true, width=0.48 \linewidth]{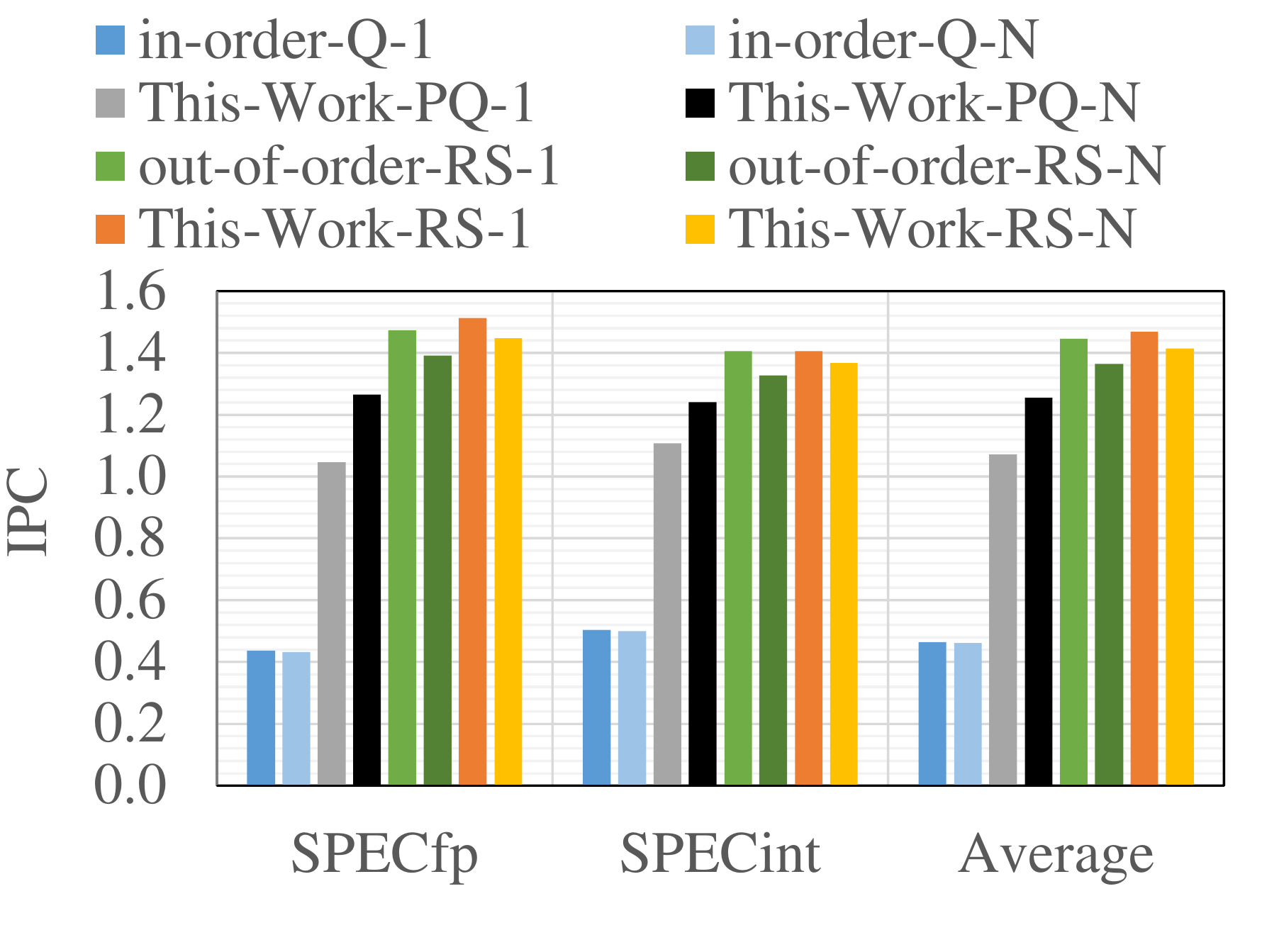}\label{fig:vliw}}
  \caption{Implementation options for (a) the prediction algorithm and (b) the execution engine (Q and PQ: issue from the head, RS: issue from any position in the queue, 1: one instruction queue for all units, N: one queue per functional unit).}
\end{figure}

Structural hazards are another major source of stalls as they can block the instruction queue by increasing resource contention of the functional units. To take into account the time an instruction waits in the instruction queue, we use the dispatch time instead of issue time in calculating the issue time prediction. However, instruction waiting times in a dynamic processor environment are unpredictable across iterations. Figure~\ref{fig:dispatch-time} shows that using the dispatch time instead of the issue time in the prediction algorithm reduces the performance by an average of 25.7\%. The dynamic nature of the proposed core changes the schedules in every iteration and the time an instruction will wait in a queue, making structural hazard delays unpredictable. 

To minimize the impact of structural hazards in an energy efficient way, the proposed core implements per functional unit priority queues (PQ-N). Figure~\ref{fig:vliw} shows results of implementing a multi-queue back-end for each baseline core. The performance of the in-order core is not improved as instructions still need to be issued in program order (in-order-Q-1 to in-order-Q-N). The proposed core (that uses a PQ-N) achieves a performance improvement of 14.7\% over a single priority queue (PQ-1) implementation. Increasing the number of reservation stations (RS) in the out-of-order processor and limiting issue of each queue to a single functional unit (RS-N), reduces its performance due to the limited destinations an instruction can issue to. Implementing an RS-1 on this work surpasses out-of-order performance by 1.6\% as the RS compensates for issue time mispredictions and removes structural hazards completely. An RS-N solution suffers from performance loss for the same reason as the out-of-order core.

\begin{figure}[!t]
  \centering
  \subfloat[\textbf{Training Frequency}]{\includegraphics[trim=2cm 7.8cm 1.5cm 7.5cm, clip=true, width=0.5 \linewidth]{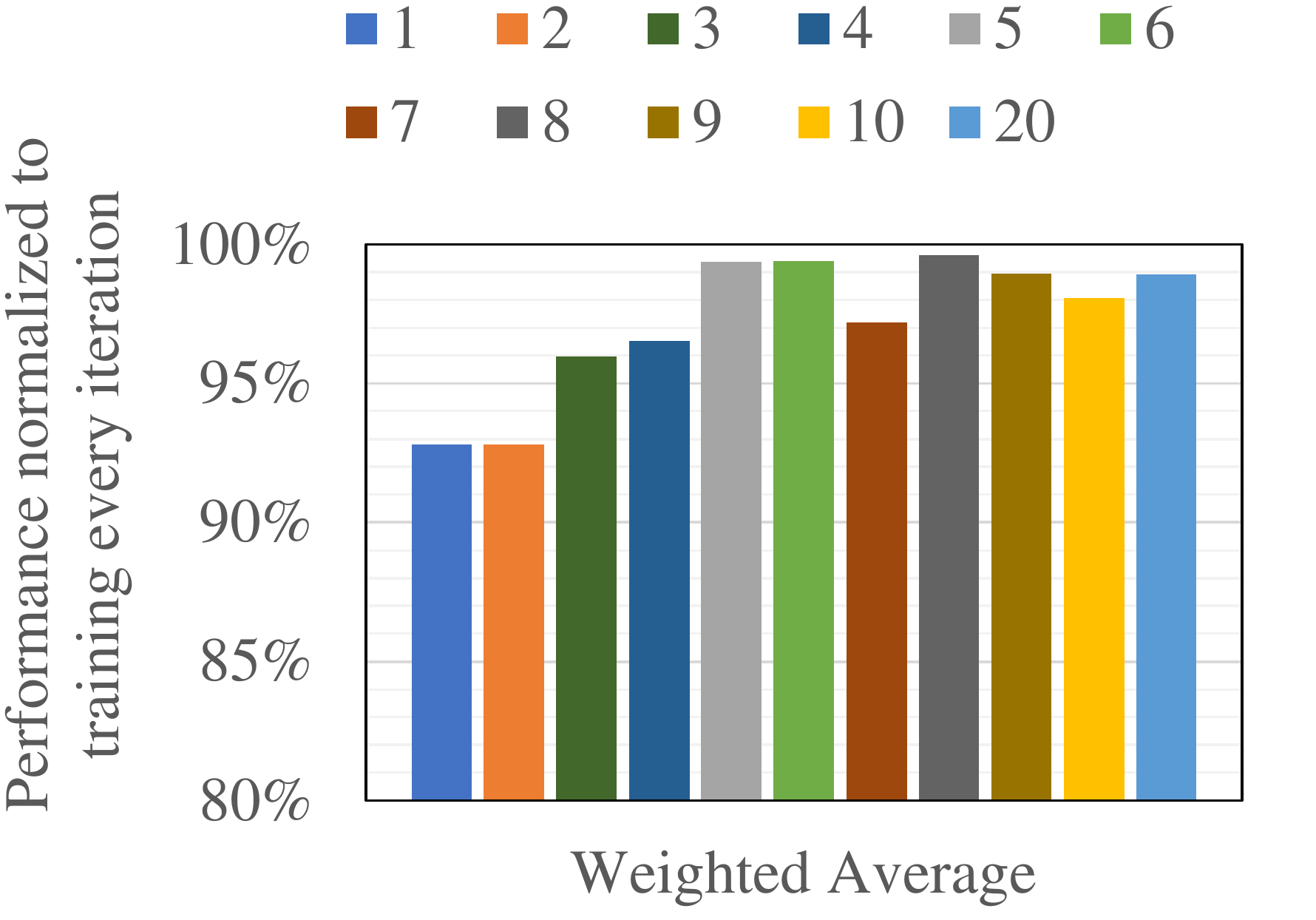}\label{fig:training}}
  \hfill
  \subfloat[\textbf{Saturating Counter}]{\includegraphics[trim=2cm 7.8cm 1.5cm 7.5cm, clip=true, width=0.5 \linewidth]{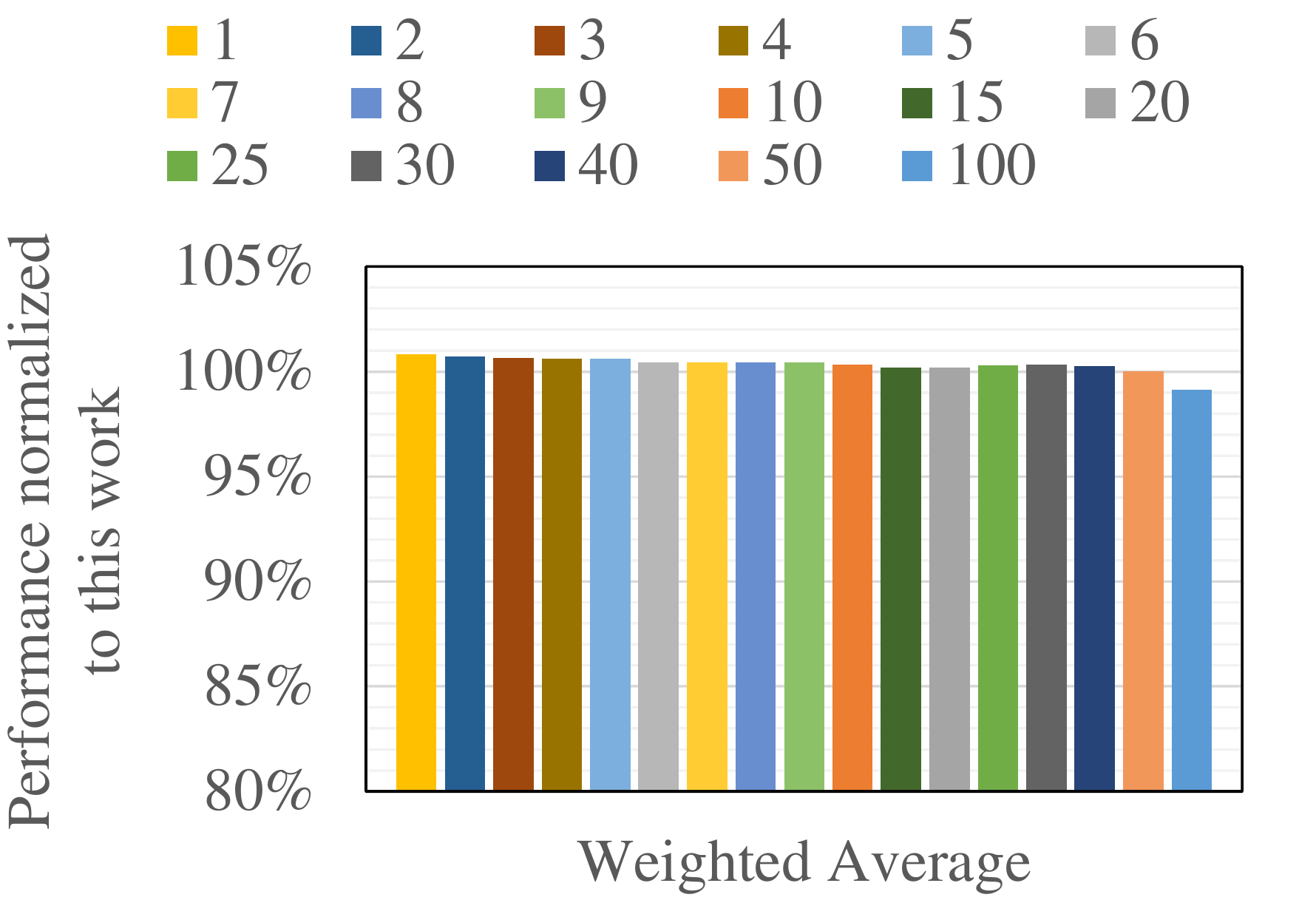}\label{fig:saturating-counter}}
  \caption{(a) Train for a number of iterations and use that prediction thereafter (normalized to training every iteration), (b) Use a saturating counter to update the delay every N identical consecutive delays (normalized to This Work). Note that the y-axis starts at 80\%.}
\end{figure}

\begin{figure}[!t]
	\centering
	\includegraphics[trim=1.5cm 9.8cm 1cm 9.5cm, clip=true, width=0.95 \linewidth]{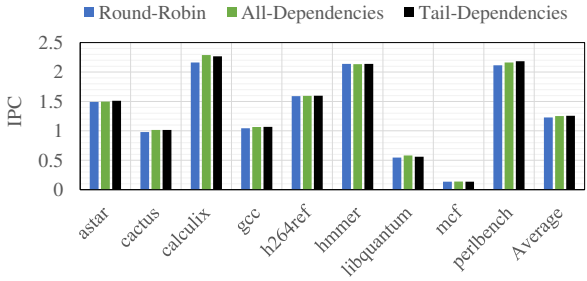}
	\caption{Steering instructions to priority queues: \textit{Round-Robin}, \textit{All-Dependencies} (follow all producers in a queue) and \textit{Tail-Dependencies} (dependent on an instruction at a queue's tail).}
	\label{fig:steering}
\end{figure}

\textbf{Issue Time Training Frequency. } Load latency analysis shows that using the previous load delay (per PC) provides an average of 92.8\% accuracy for predicting the next delay value. The training frequency of the issue time predictor depends on the application and the number of times the instruction delay changes throughout the execution. Using different predictor training frequencies (Figure~\ref{fig:training}) shows that the more often the predictor is trained, the higher the performance that can be achieved. Alternatively, using a saturating counter delay predictor that updates the delay of an instruction in the DelayCache only after the same delay appears for a number of consecutive iterations (Figure~\ref{fig:saturating-counter}) shows a marginal average improvement of 0.8\% over this work (with \texttt{sphinx3} the only exception to achieve a 10\% improvement due to the high number of consecutive misses in the cache). Both of these studies show that changes in different iterations of load access delays of repeated instructions are highly unpredictable and do not follow a specific pattern that can be easily learned. But, in-depth analysis of the delays shows that in most consecutive appearances the delay is the same (hence the 92.8\% accuracy).

A more sophisticated branch-predictor-like mechanism based on loops could store multiple delays per instruction to train the predictor with higher confidence. However, our study shows that storing as many as 5 delays per instruction and using the most frequent, the smallest, the largest or the average, in the prediction does not further improve performance.

\textbf{Instruction Steering Analysis. } As described in Section~\ref{sec:steering}, the proposed design uses dependencies on instructions at the tails of the queues to dispatch new instructions to the queues (\textit{Tail-Dependencies}). We found this technique to produce higher performance compared to either checking dependencies to all instructions in a queue (\textit{All-Dependencies}) or inserting instructions in a \textit{Round-Robin} scheme. Checking for dependencies only at the tail of each queue achieves on average 2.1\% and 0.3\% improvement over \textit{Round-Robin} and \textit{All-Dependencies} respectively (See Figure~\ref{fig:steering}).

\begin{figure}
  \centering
    \subfloat[\textbf{Priority Queue scaling}]{\includegraphics[trim=2cm 7.2cm 1.5cm 7.5cm, clip=true, width=0.48 \linewidth]{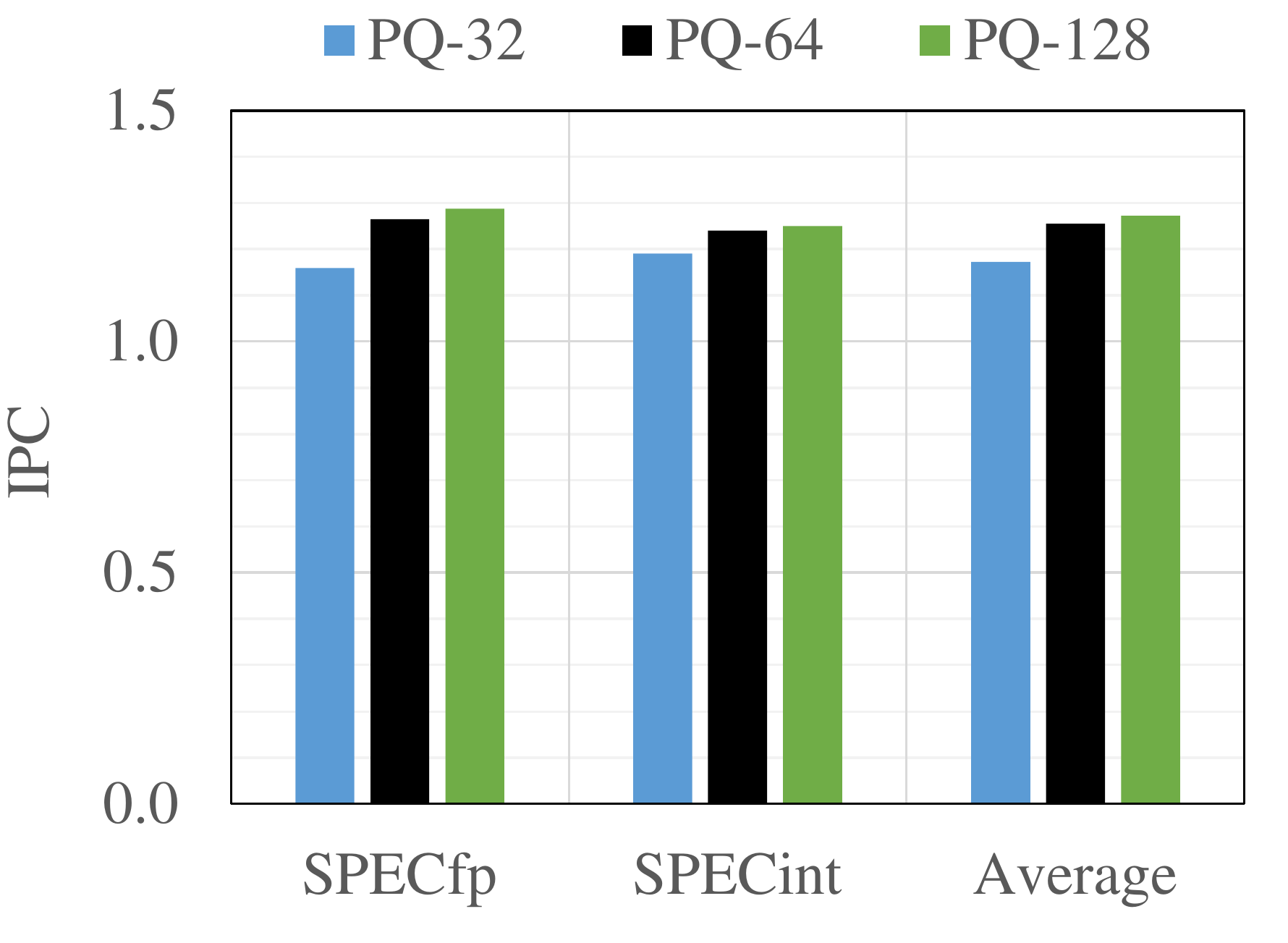}\label{fig:weak-scaling}}\hfill
    \subfloat[\textbf{Functional Units scaling}]{\includegraphics[trim=2cm 7.2cm 1.5cm 7.5cm, clip=true, width=0.48 \linewidth]{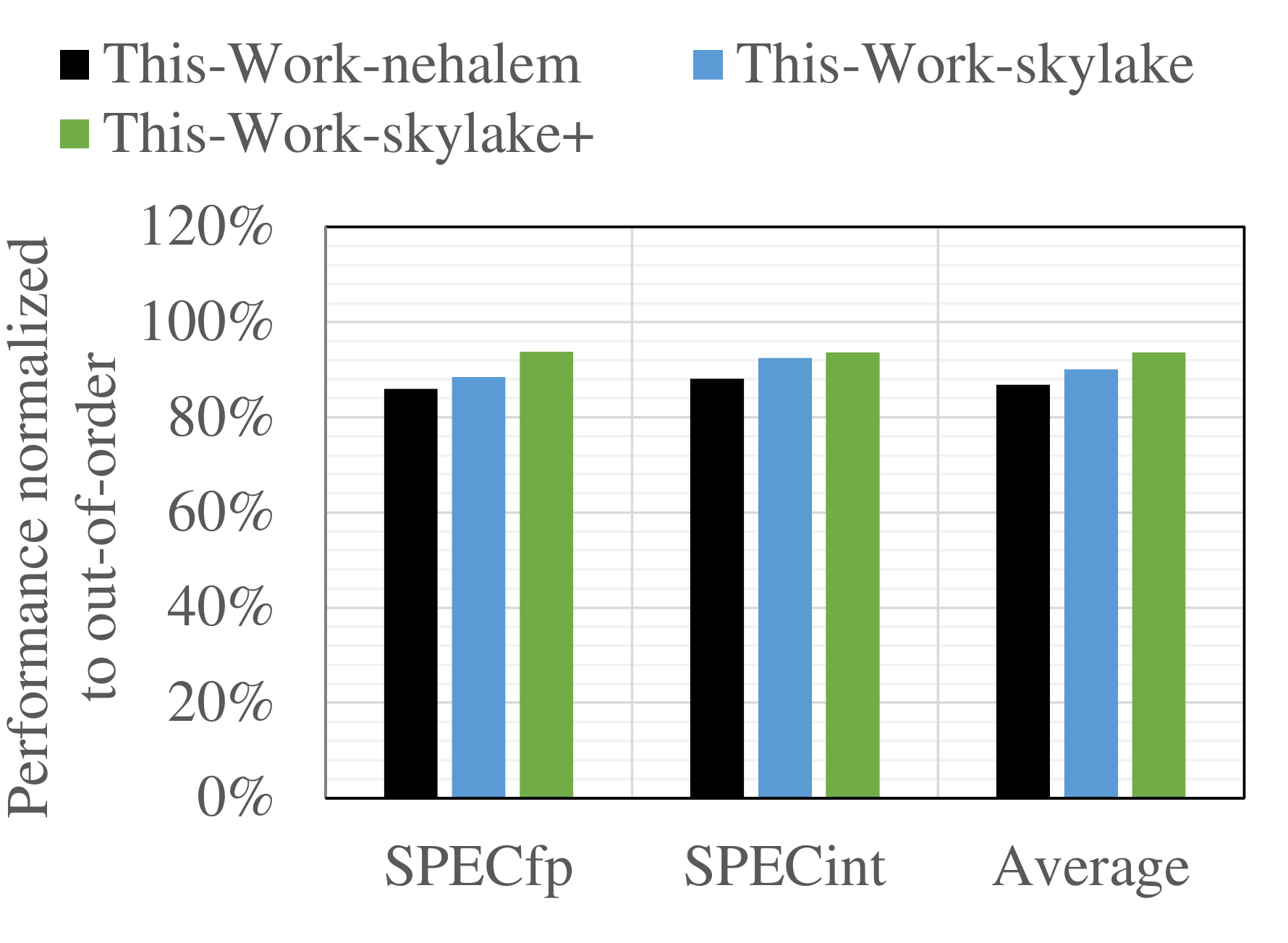}\label{fig:units-scale}}\par
    \subfloat[\textbf{DelayCache scaling}]{\includegraphics[trim=2cm 7.2cm 1.5cm 7.5cm, clip=true, width=0.48 \linewidth]{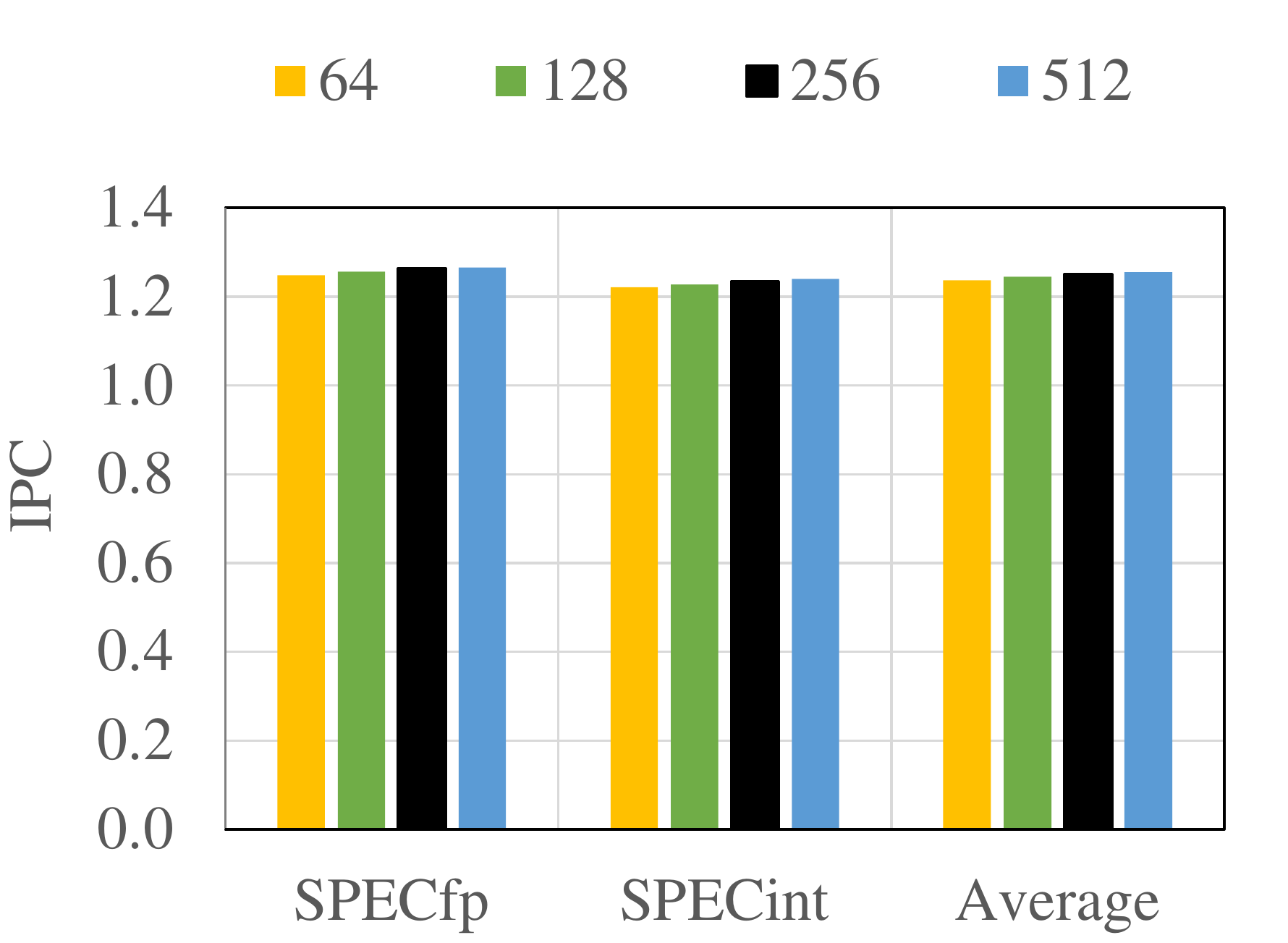}\label{fig:delayCache-scale}}
    \caption{Scaling the: (a) Priority instruction queues with a 128-entry ROB (the size refers to the total sum of all queues), %
    (b) number of functional units (each one normalized to its corresponding out-of-order baseline) and (c) DelayCache.}
  \label{fig:scaling}
\end{figure}

\textbf{Core Components Scalability. } Figure~\ref{fig:scaling} shows a scalability study of the core structures and how they affect the performance of the proposed design. The size of the priority instruction queues (Figure~\ref{fig:weak-scaling}) affects the throughput of the front-end. %
The number of functional units affects the throughput of the backend (Figure~\ref{fig:units-scale}), while the size of the DelayCache determines the number of delays that can be stored (Figure~\ref{fig:delayCache-scale}). 

Stalling of the front-end can happen when instruction queues are too small. The proposed core can stall even when a single queue is full, therefore selecting the correct size is important. Figure~\ref{fig:weak-scaling} shows that for the same ROB size (128 entries), a total of 64-entries for all queues (13 entries per queue) achieves similar performance to a total size of 128 entries. %

The port configuration of this work is based on an Intel Nehalem core that supports 3 generic, 1 load and 2 store (1 for address calculation and 1 for data) units. Figure~\ref{fig:units-scale} shows results for a Skylake-based configuration (4 generic, 2 load and 2 store units) and Skylake+ (4 generic, 2 load and 4 store units). The proposed implementation achieves performance within 10\% and within 6.4\% of their corresponding out-of-order Skylake and Skylake+ processors respectively. Figure~\ref{fig:delayCache-scale} shows that performance improvement for more than 64 entries in the DelayCache is marginal and levels off at 512-entries for all applications tested. The DelayCache can be small as we only store delays of load instructions that miss in the L1 cache.

\section{Related Work}
\label{sec:related}

\begin{scriptsize}
  \begin{table}[!t]
  \caption{State-of-the-art instruction reordering processors that try to mitigate delays coming from data-flow dependencies (\textit{Static}) and/or runtime delays (\textit{Dynamic}). \textit{Reorder} designates the stage instructions are reordered at (Back: Back-end and Front: Front-end). \textit{Scheduler} is the reordering mechanism used (RS: Reservation Station, FIFO: First In First Out, AST: Associative Table, Replay: Reschedule stalled instructions, CQ: Circular Queues and PQ: Priority Queues).}
  \label{table:motivation}
  \centering
    \resizebox{\columnwidth}{!}{%
    \begin{tabular}{|m{13em}|c|c|c|c|}
    \hline
    
    \textbf{State of the art hardware instruction reordering} &
    \rotatebox{90}{\textbf{Static}} &
    \rotatebox{90}{\textbf{Dynamic}} &
    \rotatebox{90}{\textbf{Reorder}} &
    \rotatebox{90}{\textbf{Scheduler}}\\
    \hline
    
    Data-flow Prescheduling~\cite{dataflow-prescheduling} & \checkmark & & Front \& Back & RS \\ \hline
    Wait Instruction Buffer~\cite{Lebeck} & & \checkmark & Front \& Back & RS\\ \hline
    Long-Term Parking~\cite{ltp} & \checkmark & \checkmark & Front \& Back & RS\\ \hline
    Insequence Instructions~\cite{ooo-smt} & \checkmark & & Back & RS\\ \hline
    WiDGET~\cite{widget} & \checkmark & & Back & RS\\ \hline
    Runahead~\cite{runahead} & & \checkmark & Back & RS\\ \hline
    Continuous Runahead~\cite{continuous-runahead} & & \checkmark & Back & RS\\ \hline
    Load Scheduling~\cite{miss-predict} & & \checkmark & Back & RS\\ \hline
    Segmented IQs~\cite{segmentedIQs} & \checkmark & \checkmark & Back & RS\\ \hline
    Look-ahead Prediction~\cite{scaling-issue-window} & \checkmark & \checkmark & Back & RS\\ \hline
    Dynamos~\cite{dynamos} & \checkmark & \checkmark & Back & RS+FIFO\\ \hline
    Mirage~\cite{mirage} & \checkmark & \checkmark & Back & RS+FIFO\\ \hline
    FIFOrder~\cite{fiforder} & \checkmark & \checkmark & Front \& Back & RS+FIFO\\ \hline
    Dealy and Bypass~\cite{delayBypass} & \checkmark & \checkmark & Front \& Back & RS+FIFO\\ \hline
    N-use Issue Logic~\cite{gonzalez2001} & \checkmark & \checkmark & Front \& Back & AST+FIFO\\ \hline
    Deterministic Issue Logic~\cite{gonzalez2001} & \checkmark & & Back & RS+CQ\\ \hline
    Distance Issue Logic~\cite{gonzalez2000} & \checkmark & & Back & RS+CQ\\ \hline
    In-order SMT~\cite{seznec} & \checkmark & & Front & FIFO\\ \hline
    Load Slice Core~\cite{lsc} & \checkmark &  & Front & FIFO\\ \hline
    Freeway~\cite{freeway} & \checkmark & \checkmark  & Front & FIFO\\ \hline
    Complexity-Effective~\cite{complexity-effective} & \checkmark & & Front & FIFO\\\hline
    iCFP~\cite{icfp} & & \checkmark & Front & FIFO\\ \hline
    CASINO~\cite{casino} & & \checkmark & Front & FIFO\\ \hline
    Wakeup-free~\cite{wakeup-free} & \checkmark & \checkmark & Front \& Back & Replay\\ \hline
    Cyclone~\cite{cyclone} & \checkmark & \checkmark & Front \& Back & Replay\\ \hline
    \textit{This Work} & \checkmark & \checkmark & Back & PQ\\\hline
    
    \end{tabular}
    }%
  \end{table}
\end{scriptsize}

There has been extensive work in the past on instruction reordering to reduce runtime delays and improve processor performance. Table~\ref{table:motivation} presents state-of-the-art hardware solutions in instruction reordering, the delays they try to mitigate, the stage in the processor the reordering takes place and the type of scheduler used to reorder instructions. High performance comes from mitigating both \textit{Static} and \textit{Dynamic} delays, while reordering instructions in the backend of the processor provides for higher flexibility. Unfortunately, the majority of past solutions use an \textit{RS-based} scheduler for reordering instructions which limits energy efficiency improvement. In this work we argue that smarter solutions are needed to significantly improve energy efficiency, using a simpler and more scalable scheduler (\textit{PQ}) to reorder instructions in the backend, while achieving high performance by addressing \textit{Static} and \textit{Dynamic} delays. In this section we discuss different categories of solutions that address runtime delays in instruction scheduling.

\textbf{RS-based Schedulers. } Many solutions use data-flow dependencies to preschedule or prioritize instructions in order to improve the performance or the efficiency of an out-of-order processor. Data-flow Prescheduling~\cite{dataflow-prescheduling} fetches and reorders instructions in a prescheduling buffer using data-flow dependencies. This provides for a larger effective window size while keeping the issue buffer small. However, it does not take into account variable delay instructions and assumes static delays for all instructions (all loads are presumed to hit in L1). Segmented Instruction Queues~\cite{segmentedIQs} divide large instruction queues into smaller segments which can be clocked at higher frequencies. They use dynamic dependence-based scheduling to promote instructions from segment to segment until they reach a small issue buffer. Data Cache Hit-Miss Prediction~\cite{miss-predict} tries to predict L1 hits and reschedule load dependent instructions based on that information. But predicting only L1 hits does not take into account off-chip memory delays that have the most impact on the the performance of a processor. In a more complex implementation Look-ahead Prediction~\cite{scaling-issue-window}, tries to predict load delays using a value predictor. Dynamic solutions like~\cite{Lebeck, ltp} predict and prioritize critical or independent instructions. In~\cite{Lebeck} instructions that depend on long-latency operations are moved from the issue queue to a much larger waiting instruction buffer (WIB) until their long-latency producer completes. Long Term Parking (LTP)~\cite{ltp} analyzes instructions and parks non-critical instructions from the main instruction stream to prioritize critical ones (address-generating instructions and loads). Similarly, N-Use~\cite{gonzalez2001}, uses an associative table to park non-ready instructions, Distance Issue Logic~\cite{gonzalez2000} assumes unknown load delays and parks all their consumers in an RS IQ until their operands are produced, while Deterministic Issue Logic~\cite{gonzalez2001} assumes a static delay for all loads and only parks stalled consumers to the RS IQ. FIFOrder~\cite{fiforder} and Delay and Bypass~\cite{delayBypass} use the knowledge that an OoO-core instruction scheduler offers (availability of the instructions operands) to dispatch ready instructions to FIFO queues to reduce the size and power consumption of complex instruction queues. They differ by the type of instructions to be send to the FIFO queues based on their criticality and readiness. All these solutions require additional hardware to implement and still employ a traditional out-of-order scheduler to handle the reordering and compensate for timing mispredictions of their techniques.

\textbf{FIFO-based Schedulers. } Due to their low power consumption, in-order processors are highly energy efficient. However, they achieve significantly lower performance compared to an out-of-order processor. Complexity-Effective~\cite{complexity-effective} reorders instructions based on their dependencies. Instructions that belong to the same data-flow dependency chain are directed to dedicated in-order queues, while selection logic is used to issue instructions from the head of the queues. The Load Slice Core~\cite{lsc} extends an in-order, stall-on-use core with a second in-order pipeline that allows memory accesses and address-generating instructions to bypass stalled instructions in the main pipeline. Unfortunately, these solutions do not take into account dynamic delays. This creates large gaps between load-dependent instructions in a real execution, that stall until the producing load returns from memory, thus limiting their performance improvement.

Cyclone~\cite{cyclone} uses a store set dependence predictor to monitor memory dependencies, while mispredicted instructions are replayed from the tail of the queue. But, this implementation potentially scrambles the ordering of other instructions in the instruction window, creating a performance bottleneck. Wakeup-free scheduling~\cite{wakeup-free} improves this structural constraints by using a collapsing scheme that does not allow instructions to move while their latency counters are decreasing. But their evaluation is done using a perfect L1 Hit predictor for load latency delays, that does not take into account off-chip memory delays that have the most impact on the the performance of a processor. iCFP~\cite{icfp} uses a Continual Flow Pipeline that switches to an advance execution mode when it encounters a L1 or L2 cache miss. Miss-dependent instructions are diverted into a slice buffer, un-blocking the pipeline for miss-independent instructions to execute. Although it achieves low power consumption, its performance is limited to 68\% of the performance of an out-of-order processor~\cite{icfp}. Freeway~\cite{freeway} is an orthogonal solution that implements a technique similar to LTP~\cite{ltp} on top of an in-order core and manages to improve its performance by 80\%, while we achieve 180\% increase in performance over our in-order baseline core (on the same applications). CASINO~\cite{casino} uses two in-order queues to filter instructions that block the issuing queue. However, their solution takes no real-time information into consideration when doing the filtering that can potentially lead to even more excessive delays when one of the queues is filled, depending on the applications executed.

\textbf{Heterogeneous Processors. }  Mirage Cores~\cite{mirage} and its predecessor Dynamos~\cite{dynamos} employ a full out-of-order core to produce fast out-of-order schedules that are stored in a local cache structure and executed by a number of in-order cores on the same processor. WiDGET~\cite{widget} enables dynamic customization of different combinations of small and/or powerful cores as a way to increase performance and reduce power consumption depending on the executing workload. The design complexity and cost of these solutions however, makes them inefficient as they still require the implementation of an out-of-order core to learn aggressive instruction schedules.

\textbf{Software Implementations. } Compile-time application analysis is also used to categorize and prioritize instructions by predicting the critical path of the execution~\cite{critical-path-prediction-1, critical-path-prediction-2}. Solutions with good balance between performance and energy efficiency use modified hardware equipped with the appropriate compile-time support to statically reorder instructions in advance~\cite{speculative, braid, outrider, clairvoyance, denver, duke, swoop}. But, unlike our work, these solutions require modification to the application itself and do not provide backward compatibility for deployed applications.

\textbf{Simultaneous Multi-Threading (SMT). } In a multi-threaded architecture, independent instructions from different threads can be used to overcome dependency stalls from a single thread~\cite{seznec, ooo-smt}. This boosts performance of multi-threaded applications as it increases processor throughput in throughput-sensitive parallel applications. However, these techniques do not address single-thread performance.

\textbf{Prefetching. } Prefetching attempts to minimize cache misses by executing additional instructions~\cite{runahead, data-cache-preexecution, continuous-runahead}. Runahead~\cite{runahead} allows the execution to continue past stalling to pre-execute instructions and generate new cache misses that fetch data earlier for future instructions. Continuous runahead~\cite{continuous-runahead} extends previous solutions by dynamically filtering the instruction stream to identify the chains of operations that cause a pipeline to stall. Unfortunately, prefetching techniques alone are not enough as they only try to hide memory latency. All solutions referenced here still use a complex out-of-order scheduler to handle instructions reordering.

\section{Conclusion}
\label{sec:conclusions}

In this work, we propose a novel scheduling scheme that tracks real-time delays of load instructions to accurately predict instruction issue times, and a priority-based instruction reordering mechanism that achieves near out-of-order performance in an energy efficient way. To this end, we design a new microarchitecture that builds aggressive schedules and produces near out-of-order performance in an energy efficient way. The proposed design replaces the complex instruction scheduler of an out-of-order processor with a \textit{instruction delay learning mechanism} that monitors load instructions and learns their latest real-time delays, an \textit{issue time predictor} that predicts their issue times and \textit{priority queue reordering} that efficiently reorder instructions. Together, these three techniques allow the new core to achieve 86.2\% of the performance of the baseline out-of-order, while reducing the power consumption for instruction scheduling hardware by 88\%.

\bibliographystyle{IEEEtran}
\bibliography{references}

\begin{thebibliography}{10}
\providecommand{\url}[1]{#1}
\csname url@samestyle\endcsname
\providecommand{\newblock}{\relax}
\providecommand{\bibinfo}[2]{#2}
\providecommand{\BIBentrySTDinterwordspacing}{\spaceskip=0pt\relax}
\providecommand{\BIBentryALTinterwordstretchfactor}{4}
\providecommand{\BIBentryALTinterwordspacing}{\spaceskip=\fontdimen2\font plus
\BIBentryALTinterwordstretchfactor\fontdimen3\font minus
  \fontdimen4\font\relax}
\providecommand{\BIBforeignlanguage}[2]{{%
\expandafter\ifx\csname l@#1\endcsname\relax
\typeout{** WARNING: IEEEtran.bst: No hyphenation pattern has been}%
\typeout{** loaded for the language `#1'. Using the pattern for}%
\typeout{** the default language instead.}%
\else
\language=\csname l@#1\endcsname
\fi
#2}}
\providecommand{\BIBdecl}{\relax}
\BIBdecl

\bibitem{dynamos}
\BIBentryALTinterwordspacing
S.~Padmanabha, A.~Lukefahr, R.~Das, and S.~Mahlke, ``Dynamos: Dynamic schedule
  migration for heterogeneous cores,'' in \emph{Proceedings of the 48th
  International Symposium on Microarchitecture}, ser. MICRO-48.\hskip 1em plus
  0.5em minus 0.4em\relax New York, NY, USA: ACM, 2015, pp. 322--333. [Online].
  Available: \url{http://doi.acm.org/10.1145/2830772.2830791}
\BIBentrySTDinterwordspacing

\bibitem{mirage}
\BIBentryALTinterwordspacing
S.~Padmanabha\vspace{0mm}, A.~Lukefahr, R.~Das, and S.~Mahlke, ``Mirage cores:
  The illusion of many out-of-order cores using in-order hardware,'' in
  \emph{Proceedings of the 50th Annual IEEE/ACM International Symposium on
  Microarchitecture}, ser. MICRO-50 '17.\hskip 1em plus 0.5em minus 0.4em\relax
  New York, NY, USA: ACM, 2017, pp. 745--758. [Online]. Available:
  \url{http://doi.acm.org/10.1145/3123939.3123969}
\BIBentrySTDinterwordspacing

\bibitem{ltp}
A.~Sembrant, T.~Carlson, E.~Hagersten, D.~Black-Shaffer, A.~Perais, A.~Seznec,
  and P.~Michaud, ``Long term parking (ltp): Criticality-aware resource
  allocation in ooo processors,'' in \emph{2015 48th Annual IEEE/ACM
  International Symposium on Microarchitecture (MICRO)}, Dec 2015, pp.
  334--346.

\bibitem{Lebeck}
\BIBentryALTinterwordspacing
A.~R. Lebeck, J.~Koppanalil, T.~Li, J.~Patwardhan, and E.~Rotenberg, ``A large,
  fast instruction window for tolerating cache misses,'' in \emph{Proceedings
  of the 29th Annual International Symposium on Computer Architecture}, ser.
  ISCA '02.\hskip 1em plus 0.5em minus 0.4em\relax Washington, DC, USA: IEEE
  Computer Society, 2002, pp. 59--70. [Online]. Available:
  \url{http://dl.acm.org/citation.cfm?id=545215.545223}
\BIBentrySTDinterwordspacing

\bibitem{freeway}
R.~{Kumar}, M.~{Alipour}, and D.~{Black-Schaffer}, ``Freeway: Maximizing mlp
  for slice-out-of-order execution,'' in \emph{2019 IEEE International
  Symposium on High Performance Computer Architecture (HPCA)}, 2019, pp.
  558--569.

\bibitem{lsc}
T.~E. Carlson, W.~Heirman, O.~Allam, S.~Kaxiras, and L.~Eeckhout, ``The load
  slice core microarchitecture,'' in \emph{2015 ACM/IEEE 42nd Annual
  International Symposium on Computer Architecture (ISCA)}, June 2015, pp.
  272--284.

\bibitem{casino}
I.~{Jeong}, S.~{Park}, C.~{Lee}, and W.~W. {Ro}, ``Casino core
  microarchitecture: Generating out-of-order schedules using cascaded in-order
  scheduling windows,'' in \emph{2020 IEEE International Symposium on High
  Performance Computer Architecture (HPCA)}, 2020, pp. 383--396.

\bibitem{delayBypass}
M.~{Alipour}, S.~{Kaxiras}, D.~{Black-Schaffer}, and R.~{Kumar}, ``Delay and
  bypass: Ready and criticality aware instruction scheduling in out-of-order
  processors,'' in \emph{2020 IEEE International Symposium on High Performance
  Computer Architecture (HPCA)}, 2020, pp. 424--434.

\bibitem{fiforder}
M.~{Alipour}, R.~{Kumar}, S.~{Kaxiras}, and D.~{Black-Schaffer}, ``Fiforder
  microarchitecture: Ready-aware instruction scheduling for ooo processors,''
  in \emph{2019 Design, Automation Test in Europe Conference Exhibition
  (DATE)}, March 2019, pp. 716--721.

\bibitem{cyclone}
\BIBentryALTinterwordspacing
D.~Ernst, A.~Hamel, and T.~Austin, ``Cyclone: A broadcast-free dynamic
  instruction scheduler with selective replay,'' in \emph{Proceedings of the
  30th Annual International Symposium on Computer Architecture}, ser. ISCA
  '03.\hskip 1em plus 0.5em minus 0.4em\relax New York, NY, USA: ACM, 2003, pp.
  253--263. [Online]. Available: \url{http://doi.acm.org/10.1145/859618.859647}
\BIBentrySTDinterwordspacing

\bibitem{dataflow-prescheduling}
P.~Michaud and A.~Seznec, ``Data-flow prescheduling for large instruction
  windows in out-of-order processors,'' in \emph{Proceedings HPCA Seventh
  International Symposium on High-Performance Computer Architecture}, Jan 2001,
  pp. 27--36.

\bibitem{complexity-effective}
\BIBentryALTinterwordspacing
S.~Palacharla, N.~P. Jouppi, and J.~E. Smith, ``Complexity-effective
  superscalar processors,'' \emph{SIGARCH Comput. Archit. News}, vol.~25,
  no.~2, pp. 206--218, May 1997. [Online]. Available:
  \url{http://doi.acm.org/10.1145/384286.264201}
\BIBentrySTDinterwordspacing

\bibitem{icfp}
A.~Hilton, S.~Nagarakatte, and A.~Roth, ``{iCFP}: Tolerating all-level cache
  misses in in-order processors,'' in \emph{2009 IEEE 15th International
  Symposium on High Performance Computer Architecture}, Feb 2009, pp. 431--442.

\bibitem{wakeup-free}
J.~S. {Hu}, N.~{Vijaykrishnan}, and M.~J. {Irwin}, ``Exploring wakeup-free
  instruction scheduling,'' in \emph{10th International Symposium on High
  Performance Computer Architecture (HPCA'04)}, Feb 2004, pp. 232--232.

\bibitem{gonzalez2000}
\BIBentryALTinterwordspacing
R.~Canal and A.~Gonz\'{a}lez, ``A low-complexity issue logic,'' in
  \emph{Proceedings of the 14th International Conference on Supercomputing},
  ser. ICS ’00.\hskip 1em plus 0.5em minus 0.4em\relax New York, NY, USA:
  Association for Computing Machinery, 2000, p. 327–335. [Online]. Available:
  \url{https://doi.org/10.1145/335231.335263}
\BIBentrySTDinterwordspacing

\bibitem{gonzalez2001}
\BIBentryALTinterwordspacing
------, ``Reducing the complexity of the issue logic,'' in \emph{Proceedings of
  the 15th International Conference on Supercomputing}, ser. ICS ’01.\hskip
  1em plus 0.5em minus 0.4em\relax New York, NY, USA: Association for Computing
  Machinery, 2001, p. 312–320. [Online]. Available:
  \url{https://doi.org/10.1145/377792.377854}
\BIBentrySTDinterwordspacing

\bibitem{widget}
\BIBentryALTinterwordspacing
Y.~Watanabe, J.~D. Davis, and D.~A. Wood, ``{WiDGET}: Wisconsin decoupled grid
  execution tiles,'' \emph{SIGARCH Comput. Archit. News}, vol.~38, no.~3, pp.
  2--13, Jun. 2010. [Online]. Available:
  \url{http://doi.acm.org/10.1145/1816038.1815965}
\BIBentrySTDinterwordspacing

\bibitem{ooo-smt}
F.~M. Sleiman and T.~F. Wenisch, ``Efficiently scaling out-of-order cores for
  simultaneous multithreading,'' in \emph{2016 ACM/IEEE 43rd Annual
  International Symposium on Computer Architecture (ISCA)}, June 2016, pp.
  431--443.

\bibitem{scaling-issue-window}
\BIBentryALTinterwordspacing
Y.~Liu, A.~Shayesteh, G.~Memik, and G.~Reinman, ``Scaling the issue window with
  look-ahead latency prediction,'' in \emph{Proceedings of the 18th Annual
  International Conference on Supercomputing}, ser. ICS '04.\hskip 1em plus
  0.5em minus 0.4em\relax New York, NY, USA: ACM, 2004, pp. 217--226. [Online].
  Available: \url{http://doi.acm.org/10.1145/1006209.1006240}
\BIBentrySTDinterwordspacing

\bibitem{segmentedIQs}
S.~E. {Raasch}, N.~L. {Binkert}, and S.~K. {Reinhardt}, ``A scalable
  instruction queue design using dependence chains,'' in \emph{Proceedings 29th
  Annual International Symposium on Computer Architecture}, May 2002, pp.
  318--329.

\bibitem{miss-predict}
\BIBentryALTinterwordspacing
A.~Yoaz, M.~Erez, R.~Ronen, and S.~Jourdan, ``Speculation techniques for
  improving load related instruction scheduling,'' in \emph{Proceedings of the
  26th Annual International Symposium on Computer Architecture}, ser. ISCA
  '99.\hskip 1em plus 0.5em minus 0.4em\relax Washington, DC, USA: IEEE
  Computer Society, 1999, pp. 42--53. [Online]. Available:
  \url{http://dx.doi.org/10.1145/300979.300983}
\BIBentrySTDinterwordspacing

\bibitem{asplos2013}
\BIBentryALTinterwordspacing
D.~S. McFarlin, C.~Tucker, and C.~Zilles, ``Discerning the dominant
  out-of-order performance advantage: Is it speculation or dynamism?'' in
  \emph{Proceedings of the Eighteenth International Conference on Architectural
  Support for Programming Languages and Operating Systems}, ser. ASPLOS
  '13.\hskip 1em plus 0.5em minus 0.4em\relax New York, NY, USA: ACM, 2013, pp.
  241--252. [Online]. Available:
  \url{http://doi.acm.org/10.1145/2451116.2451143}
\BIBentrySTDinterwordspacing

\bibitem{leiserson1979systolic}
C.~E. Leiserson, ``Systolic priority queues.'' CARNEGIE-MELLON UNIV PITTSBURGH
  PA DEPT OF COMPUTER SCIENCE, Tech. Rep., 1979.

\bibitem{store-set-predictor}
\BIBentryALTinterwordspacing
G.~Z. Chrysos and J.~S. Emer, ``Memory dependence prediction using store
  sets,'' \emph{SIGARCH Comput. Archit. News}, vol.~26, no.~3, p. 142–153,
  Apr. 1998. [Online]. Available: \url{https://doi.org/10.1145/279361.279378}
\BIBentrySTDinterwordspacing

\bibitem{sniper}
T.~E. {Carlson}, W.~{Heirman}, and L.~{Eeckhout}, ``Sniper: Exploring the level
  of abstraction for scalable and accurate parallel multi-core simulation,'' in
  \emph{SC '11: Proceedings of 2011 International Conference for High
  Performance Computing, Networking, Storage and Analysis}, Nov 2011, pp.
  1--12.

\bibitem{sniper-taco}
\BIBentryALTinterwordspacing
T.~E. Carlson, W.~Heirman, S.~Eyerman, I.~Hur, and L.~Eeckhout, ``An evaluation
  of high-level mechanistic core models,'' \emph{ACM Trans. Archit. Code
  Optim.}, vol.~11, no.~3, Aug. 2014. [Online]. Available:
  \url{https://doi.org/10.1145/2629677}
\BIBentrySTDinterwordspacing

\bibitem{mcpat}
\BIBentryALTinterwordspacing
S.~Li, J.~H. Ahn, R.~D. Strong, J.~B. Brockman, D.~M. Tullsen, and N.~P.
  Jouppi, ``Mcpat: An integrated power, area, and timing modeling framework for
  multicore and manycore architectures,'' in \emph{Proceedings of the 42Nd
  Annual IEEE/ACM International Symposium on Microarchitecture}, ser. MICRO
  42.\hskip 1em plus 0.5em minus 0.4em\relax New York, NY, USA: ACM, 2009, pp.
  469--480. [Online]. Available:
  \url{http://doi.acm.org/10.1145/1669112.1669172}
\BIBentrySTDinterwordspacing

\bibitem{simpoint}
\BIBentryALTinterwordspacing
T.~Sherwood, E.~Perelman, G.~Hamerly, and B.~Calder, ``Automatically
  characterizing large scale program behavior,'' in \emph{Proceedings of the
  10th International Conference on Architectural Support for Programming
  Languages and Operating Systems}, ser. ASPLOS X.\hskip 1em plus 0.5em minus
  0.4em\relax New York, NY, USA: ACM, 2002, pp. 45--57. [Online]. Available:
  \url{http://doi.acm.org/10.1145/605397.605403}
\BIBentrySTDinterwordspacing

\bibitem{seznec:hal-01086920}
\BIBentryALTinterwordspacing
A.~Seznec, ``{TAGE-SC-L Branch Predictors},'' in \emph{{JILP - Championship
  Branch Prediction}}, Minneapolis, United States, Jun. 2014. [Online].
  Available: \url{https://hal.inria.fr/hal-01086920}
\BIBentrySTDinterwordspacing

\bibitem{power}
S.~Mishra, N.~Singh, and V.~Rousseau, \emph{Understanding Power Consumption
  Fundamentals}, 12 2016, pp. 13--27.

\bibitem{runahead}
O.~Mutlu, J.~Stark, C.~Wilkerson, and Y.~N. Patt, ``Runahead execution: an
  alternative to very large instruction windows for out-of-order processors,''
  in \emph{The Ninth International Symposium on High-Performance Computer
  Architecture, 2003. HPCA-9 2003. Proceedings.}, Feb 2003, pp. 129--140.

\bibitem{continuous-runahead}
\BIBentryALTinterwordspacing
M.~Hashemi, O.~Mutlu, and Y.~N. Patt, ``Continuous runahead: Transparent
  hardware acceleration for memory intensive workloads,'' in \emph{The 49th
  Annual IEEE/ACM International Symposium on Microarchitecture}, ser.
  MICRO-49.\hskip 1em plus 0.5em minus 0.4em\relax Piscataway, NJ, USA: IEEE
  Press, 2016, pp. 61:1--61:12. [Online]. Available:
  \url{http://dl.acm.org/citation.cfm?id=3195638.3195712}
\BIBentrySTDinterwordspacing

\bibitem{seznec}
S.~Hily and A.~Seznec, ``Out-of-order execution may not be cost-effective on
  processors featuring simultaneous multithreading,'' in \emph{Proceedings
  Fifth International Symposium on High-Performance Computer Architecture}, Jan
  1999, pp. 64--67.

\bibitem{critical-path-prediction-1}
E.~Tune, D.~Liang, D.~M. Tullsen, and B.~Calder, ``Dynamic prediction of
  critical path instructions,'' in \emph{Proceedings HPCA Seventh International
  Symposium on High-Performance Computer Architecture}, Jan 2001, pp. 185--195.

\bibitem{critical-path-prediction-2}
\BIBentryALTinterwordspacing
B.~Fields, S.~Rubin, and R.~Bod\'{\i}k, ``Focusing processor policies via
  critical-path prediction,'' \emph{SIGARCH Comput. Archit. News}, vol.~29,
  no.~2, pp. 74--85, May 2001. [Online]. Available:
  \url{http://doi.acm.org/10.1145/384285.379253}
\BIBentrySTDinterwordspacing

\bibitem{speculative}
\BIBentryALTinterwordspacing
C.~Zilles and G.~Sohi, ``Execution-based prediction using speculative slices,''
  in \emph{Proceedings of the 28th Annual International Symposium on Computer
  Architecture}, ser. ISCA'01.\hskip 1em plus 0.5em minus 0.4em\relax New York,
  NY, USA: ACM, 2001, pp. 2--13. [Online]. Available:
  \url{http://doi.acm.org/10.1145/379240.379246}
\BIBentrySTDinterwordspacing

\bibitem{braid}
\BIBentryALTinterwordspacing
F.~Tseng and Y.~N. Patt, ``Achieving out-of-order performance with almost
  in-order complexity,'' in \emph{Proceedings of the 35th Annual International
  Symposium on Computer Architecture}, ser. ISCA '08.\hskip 1em plus 0.5em
  minus 0.4em\relax Washington, DC, USA: IEEE Computer Society, 2008, pp.
  3--12. [Online]. Available: \url{https://doi.org/10.1109/ISCA.2008.23}
\BIBentrySTDinterwordspacing

\bibitem{outrider}
\BIBentryALTinterwordspacing
N.~C. Crago and S.~J. Patel, ``Outrider: Efficient memory latency tolerance
  with decoupled strands,'' in \emph{Proceedings of the 38th Annual
  International Symposium on Computer Architecture}, ser. ISCA '11.\hskip 1em
  plus 0.5em minus 0.4em\relax New York, NY, USA: ACM, 2011, pp. 117--128.
  [Online]. Available: \url{http://doi.acm.org/10.1145/2000064.2000079}
\BIBentrySTDinterwordspacing

\bibitem{clairvoyance}
K.~Tran, T.~E. Carlson, K.~Koukos, M.~Själander, V.~Spiliopoulos, S.~Kaxiras,
  and A.~Jimborean, ``Clairvoyance: Look-ahead compile-time scheduling,'' in
  \emph{2017 IEEE/ACM International Symposium on Code Generation and
  Optimization (CGO)}, Feb 2017, pp. 171--184.

\bibitem{denver}
D.~Boggs, G.~Brown, N.~Tuck, and K.~S. Venkatraman, ``Denver: Nvidia's first
  64-bit arm processor,'' \emph{IEEE Micro}, vol.~35, no.~2, pp. 46--55, Mar
  2015.

\bibitem{duke}
\BIBentryALTinterwordspacing
Z.~Huang, A.~D. Hilton, and B.~C. Lee, ``Decoupling loads for nano-instruction
  set computers,'' in \emph{Proceedings of the 43rd International Symposium on
  Computer Architecture}, ser. ISCA '16.\hskip 1em plus 0.5em minus 0.4em\relax
  Piscataway, NJ, USA: IEEE Press, 2016, pp. 406--417. [Online]. Available:
  \url{https://doi.org/10.1109/ISCA.2016.43}
\BIBentrySTDinterwordspacing

\bibitem{swoop}
\BIBentryALTinterwordspacing
K.-A. Tran, A.~Jimborean, T.~E. Carlson, K.~Koukos, M.~Sj\"{a}lander, and
  S.~Kaxiras, ``Swoop: Software-hardware co-design for non-speculative,
  execute-ahead, in-order cores,'' in \emph{Proceedings of the 39th ACM SIGPLAN
  Conference on Programming Language Design and Implementation}, ser. PLDI
  2018.\hskip 1em plus 0.5em minus 0.4em\relax New York, NY, USA: ACM, 2018,
  pp. 328--343. [Online]. Available:
  \url{http://doi.acm.org/10.1145/3192366.3192393}
\BIBentrySTDinterwordspacing

\bibitem{data-cache-preexecution}
\BIBentryALTinterwordspacing
J.~Dundas and T.~Mudge, ``Improving data cache performance by pre-executing
  instructions under a cache miss,'' in \emph{Proceedings of the 11th
  International Conference on Supercomputing}, ser. ICS '97.\hskip 1em plus
  0.5em minus 0.4em\relax New York, NY, USA: ACM, 1997, pp. 68--75. [Online].
  Available: \url{http://doi.acm.org/10.1145/263580.263597}
\BIBentrySTDinterwordspacing

\end{thebibliography}

\end{document}